%% file: paper.tex
\pdfoutput=1
\documentclass[a4paper,11pt]{article}
\usepackage{jheppub}    %
\usepackage{booktabs,multirow,array}
\usepackage[font=small,labelfont=bf,labelsep=period]{caption}
\providecommand{\doi}[1]{doi:#1}   %
\usepackage{xcolor}
\graphicspath{{paper/figures/}}

\makeatletter
\renewcommand{\@fpheader}{}   %
\makeatother

\input{notes/notation}

\title{First Measurement of Bottom Flux in $e^+e^-$ Collisions at $\sqrt{s}=91.2$~GeV with DELPHI Open Data}

\author[a]{Sanghyun Ko,}
\author[b]{Kyle Lee,}
\author[c] {Dietrich Liko,}
\author[d]{Ian Moult,}
\author[e]{Sitian Qian}
\author[f]{and Jingyu Zhang}

\affiliation[a]{CERN, 1211 Geneva 23, Switzerland}
\affiliation[b]{High Energy Physics Division, Argonne National Laboratory, Lemont, Illinois 60439, USA}
\affiliation[c]{Marietta Blau Institute for Particle Physics, Austrian Academy of Sciences, 1010 Vienna, Austria}
\affiliation[d]{Department of Physics, Yale University, New Haven, Connecticut 06511, USA}
\affiliation[e]{Department of Physics and Astronomy, Northwestern University, Evanston, Illinois 60208, USA}
\affiliation[f]{Department of Physics and Astronomy, Vanderbilt University, Nashville, Tennessee 37235, USA}

\abstract{The forward-backward asymmetries of quarks at the $Z$ pole are key observables of the electroweak precision program. They have recently been formulated as fluxes of conserved charges, which require no reconstruction of quark directions, jet, or hemisphere axes. The bottom flux is the angular distribution of the bottom charge carried by hadrons, and it determines the forward-backward asymmetry of $b$ hadrons, which is related to the $b$-quark asymmetry through calculable QCD and electroweak corrections. We present the first measurement of the bottom flux, using the DELPHI open data recorded in 1994 at $\sqrt{s}=91.2$~GeV. With a simple cut-based tagger, calibrated in situ on the data themselves, we measure the forward-backward asymmetry of $b$ hadrons with a relative precision of 10.6\%. We discuss the flavor-tagging requirements of such measurements and motivate future precision measurements with the LEP and SLC data and at future colliders. This analysis is performed with a new software framework that converts the DELPHI open data to the EDM4hep event data model.}

\begin{document}
\maketitle
\flushbottom

\section{Introduction}
\label{sec:intro}

The forward-backward asymmetries of fermions at the $Z$ pole provide valuable measurements of the chiral structure of the $Z$ couplings and one of the most precise determinations of the weak mixing angle. The pole asymmetry of $b$ quarks, $\AFBz$, is the most precise of the quark asymmetries and the single most precise determination of the weak mixing angle at LEP. The combined LEP measurement shows the largest tension of any input to the Standard Model fit, 2.8 standard deviations in the final LEP and SLD combination~\cite{ALEPH:2005ab} and 2.2 to 2.3 with the current inputs~\cite{ParticleDataGroup:2026mpi,Fischer:2026bka}. The weak mixing angle derived from it differs by 3.2 standard deviations from the one derived from the left-right asymmetry measured at SLD. Models in which new physics couples mostly to the third generation, such as vector-like quarks that mix with the $b$ quark~\cite{Choudhury:2001hs} or composite Higgs~\cite{Agashe:2006at}, shift the $Z$ couplings of the $b$ quark and can account for the tension. Whether it is a fluctuation, a systematic effect of the method, or a sign of new physics can only be decided by an independent measurement.

The LEP measurements of $\AFBz$~\cite{ALEPH:2001mdb,ALEPH:2001pzx,DELPHI:2004wzo,DELPHI:2003fml,L3:1998bss,L3:1998jgx,OPAL:2002vqi,OPAL:2003pfe} shared one construction. The thrust axis divided the event into two hemispheres and estimated the direction of the quark pair. To decide which hemisphere held the $b$ quark, the LEP analyses used the charge of a lepton, the charge of a secondary vertex, or the jet charge~\cite{Field:1977fa}, a momentum-weighted sum of the track charges in the hemisphere. The main difficulty in interpreting these measurements lies in the definition of the observable. The measured asymmetry is corrected for gluon radiation to $\AFBz$ by a QCD factor computed to second order in $\alpha_s$~\cite{Altarelli:1992fs,Ravindran:1998jw,Catani:1999nf}, defined with respect to a parton-level thrust axis signed by the direction of the $b$ quark. The correction therefore depends on how the asymmetry is defined~\cite{Weinzierl:2006yt,dEnterria:2018jsx}, and the experimental selection introduces a bias in the event topology, so the fraction of the correction that applies to a given analysis, the bias factor, was calculated with the simulation~\cite{LEPHeavyFlavorWorkingGroup:1998roh,ALEPH:2005ab}. Improved theory has thus had no direct way to meet these measurements, leaving the longstanding tension in the electroweak fit without an independent resolution.

Recent developments in theory and experiment have shown that forward-backward asymmetries can be formulated as the flux of a conserved charge~\cite{Zhang:2026emt,Lee:2026bcq}, a formulation with several theoretical and experimental advantages. The flux is infrared and collinear safe~\cite{Riembau:2024tom} and defined on the hadrons alone, without a quark, jet, or hemisphere axis, and its QCD corrections start at $\mathcal O(\alpha_s^2)$, with a coefficient proportional to the $\beta$ function. The observable has been established experimentally with a first measurement of the electric-charge flux in the DELPHI data~\cite{Zhang:2026emt,Zhang:2026vdo,Zhang:2026esd}.

Formulating the $b$-quark asymmetry $\AFBz$ as the bottom flux sharpens these advantages and suppresses the hadronization input that remains. For the electric charge, which light hadrons carry, hadronization enters through a non-perturbative correction, which the charge-flux measurement controlled with a sum rule from charge conservation. The bottom flux $\Fl$ counts the $b$ hadrons in each direction, each weighted by its bottom charge, $\pm1$~\cite{Lee:2026bcq}. Because the bottom charge is a gapped quantum number, carried only by hadrons with a mass far above the QCD scale, hadronization is suppressed and enters only at order $\LQCD/m_b$. The QCD corrections change only the amplitude of the flux, the forward-backward asymmetry of the $b$ hadrons, $A$, which is related to $\AFBz$ by a factor that is calculated to high precision and does not depend on the event topology or on the hemisphere definition of an analysis. The bottom flux is therefore a natural observable for a precision determination of $\AFBz$, on the existing data and at a future $Z$ factory.

The LEP and SLC data remain the largest sample of $Z$ decays ever recorded, and the only place where the quark asymmetries can be measured until a future $Z$ factory runs. Archived LEP and SLC data have indeed drawn new interest, both for revisiting the topics of the LEP physics program and for the planning of future colliders. The ALEPH data were analyzed first, for particle correlations, event shapes, jets, and energy correlators, and recently with flavor tagging based on deep learning~\cite{Electron-PositronAlliance:2019cpi,Electron-PositronAlliance:2021kig,Electron-PositronAlliance:2023klx, Electron-PositronAlliance:2025hze,Electron-PositronAlliance:2025fhk,Defranchis:2026wyw}. In 2024 DELPHI released its complete data, simulation, and software on the CERN Open Data portal~\cite{DELPHI:OD:release2024,DELPHI:OD:policy}. The first analyses of these data measured thrust, energy correlators~\cite{Zhang:2025nlf}, and the electric-charge flux~\cite{Zhang:2026emt,Zhang:2026vdo,Zhang:2026esd}, which opened a path from the open data to electroweak physics with the new observable. Precision on this path has two requirements: the calibration of the $b$-hadron tagger on the data, and the calibration of the detector response, in particular of the charged-track momentum scale. At a Tera-$Z$ factory such as FCC-ee~\cite{FCC:2018evy}, the statistical uncertainty on $\AFBz$ will shrink by orders of magnitude, and what limits the measurement will be these same two calibrations.

In this paper we present the first measurement of the bottom flux using the DELPHI data recorded in 1994 at $\sqrt{s}=91.2$~GeV. We develop the methodology for both precision requirements. The in-situ calibration of the tagger measures the tagging efficiency and the charge-sign dilution factor directly on the data. In parallel, we introduce a common framework for the DELPHI open data that converts both data and simulation to EDM4hep, the event data model of future-collider projects~\cite{Gaede:2022leb,Carceller:2025fjc,Fanini:2026mpx}, and distributes detector corrections, beginning with the track momentum scale, as a portable correction library. Section~\ref{sec:observable} defines the observable, and Sec.~\ref{sec:principle} describes the measurement principle. Section~\ref{sec:edm4hep} describes the data, their conversion to EDM4hep, and the momentum-scale correction, and Sec.~\ref{sec:analysis} the analysis. Section~\ref{sec:results} gives the result, and Sec.~\ref{sec:summary} a summary and an outlook. Appendix~\ref{app:validation} collects the validation of the converted data.

\section{Observable Definition}
\label{sec:observable}

The bottom charge is conserved by the strong and electromagnetic interactions. Its current is the number current of the $b$ quark, $j^\mu_{\Bn}=\bar b\gamma^\mu b$, and the bottom-flux operator,
\begin{equation}
\Qdet(\vec n)=\lim_{r\to\infty} r^2\!\int_0^\infty\!dt\; n_i\, j^i_{\Bn}(t,r\vec n),
\label{eq:Qdetop}
\end{equation}
measures the bottom charge that flows out of the collision in the direction of the unit vector $\vec n$. It is the charge flux operator of Ref.~\cite{Hofman:2008ar}, the analog of the energy flux operator, for the bottom-charge current, and the counterpart of the electric-charge flux of Refs.~\cite{Zhang:2026emt,Lee:2026bcq}. Acting on a hadronic final state $|X\rangle$, it counts the bottom charge of the hadrons at their directions,
\begin{equation}
\Qdet(\vec n)\,|X\rangle=\sum_{h\in X}\Bn_h\,\delta^{2}(\vec n-\vec n_h)\,|X\rangle ,
\label{eq:Qdet}
\end{equation}
where $\vec n_h$ is the direction of hadron $h$. We count the bottom charge with the sign of the $b$ quark, $\Bn(b)=+1$, which is opposite to the PDG convention for bottomness: $\Bn(B^-)=\Bn(\bar B^0)=\Bn(\bar B^0_s)=\Bn(\Lambda_b)=+1$ and $\Bn(B^+)=\Bn(B^0)=\Bn(B^0_s)=\Bn(\bar\Lambda_b)=-1$. Only the weak interaction changes the bottom charge, through the decay of the $b$ hadron and, for the neutral mesons, through oscillation, both on time scales far longer than hadronization. Equation~\eqref{eq:Qdet} is defined with the $b$ hadrons stable, so that $\Bn_h$ is the bottom charge at production, before any oscillation or decay. Experimentally, it is the decay that makes the hadron visible, through its displaced vertex, and the decay products carry the sign (Sec.~\ref{sec:principle}).

The expectation value of the bottom-flux operator is the one-point flavor flux correlator of the $b$ quark~\cite{Lee:2026bcq}, the bottom flux. It is a cross section weighted by the bottom charge: every event contributes the bottom charge of its hadrons at their directions,
\begin{equation}
\langle\Qdet(\vec n)\rangle=\frac{1}{\stot}\sum_X\int d\sigma_X\sum_{h\in X}\Bn_h\,\delta^{2}(\vec n-\vec n_h),
\label{eq:expval}
\end{equation}
normalized to the total hadronic cross section $\stot$. Only $Z\to b\bar b$ events contribute at leading order. In events of other flavors, $b$ hadrons arise only from $g\to b\bar b$, in pairs whose contributions cancel on average. For unpolarized beams, the expectation value depends only on the polar angle $\theta$ of $\vec n$ to the $e^-$ beam. We write it per $b\bar b$ event and per unit $c\equiv\cos\theta$, together with the unsigned density of $b$ hadrons,
\begin{equation}
n(\theta)\equiv\frac{1}{\Nbb}\frac{dN_B}{dc},\qquad
\Fl(\theta)\equiv\frac{1}{\Nbb}\frac{d}{dc}\sum_h\Bn_h
=\frac{2\pi}{R_b}\,\langle\Qdet(\vec n)\rangle ,
\qquad
\int_{-1}^{1}n\,dc=2 ,
\label{eq:nF}
\end{equation}
where $N_B$ is the number of $b$ hadrons and $R_b=\sbb/\stot$.

The shape of the flux follows from the cross section. At Born level, for $Z\to b\bar b$ with unpolarized beams,
\begin{equation}
\frac{1}{\sbbz}\frac{d\sigma^{(0)}}{dc_q}
 = \frac{3}{8}\,(1+c_q^2) + \AFBz\, c_q ,
\qquad
\AFBz = \frac{3}{4}\,\Ae\,\Ab ,
\qquad
\mathcal{A}_f = \frac{2 g_V^f g_A^f}{(g_V^f)^2+(g_A^f)^2} ,
\label{eq:born}
\end{equation}
where $\sigma^{(0)}$ and $\sbbz$ are the Born-level differential and total cross sections of $Z\to b\bar b$, $c_q=\cos\theta_q$ with $\theta_q$ the angle between the $b$ quark and the $e^-$ beam, $\mathcal A_f$ is the coupling asymmetry of fermion $f$, with $\Ae$ and $\Ab$ its values for the electron and the $b$ quark, and $g_{V,A}^f$ are the vector and axial-vector couplings of $f$ to the $Z$. The two terms have opposite symmetry under $c_q\to-c_q$. The even term carries the rate; the odd term carries all of the parity violation, and its coefficient is the forward-backward asymmetry. The $b$ hadron ($\Bn=+1$) goes along the $b$ quark and the $\bar b$ hadron ($\Bn=-1$) opposite to it, so the even term cancels between the two, and the Born-level flux and unsigned density, $\Fl^{(0)}$ and $n^{(0)}$, keep only the odd and the even term, respectively:
\begin{equation}
\Fl^{(0)}(\theta)=2\,\AFBz\,c ,
\qquad
n^{(0)}(\theta)=\tfrac34\,(1+c^2).
\label{eq:bornflux}
\end{equation}

This cancellation is exact beyond Born level and at hadron level. For unpolarized beams the angular distribution of a hadron $h$ has three parts, to all orders in QCD,
\begin{equation}
\frac{1}{\stot}\frac{d\sigma_h}{dc}
=\frac38\,(1+c^2)\,\sigma^h_T+\frac34\,(1-c^2)\,\sigma^h_L+\frac34\,c\,\sigma^h_A ,
\label{eq:TLA}
\end{equation}
a transverse and a longitudinal part, even in $c$, and a parity-odd part, which is linear in $c$ because a single spin-1 boson is exchanged; $\sigma^h_{T,L,A}$ are the corresponding cross sections for hadron $h$, normalized to the hadronic cross section~\cite{ParticleDataGroup:2026mpi}. Invariance under $CP$, which holds for the $Z$ exchange and for QCD, maps a hadron at $c$ onto its antiparticle at $-c$: $\sigma^{\bar h}_{T,L}=\sigma^h_{T,L}$ and $\sigma^{\bar h}_A=-\sigma^h_A$. Weighting by $\Bn_h=\pm1$ therefore removes the even parts exactly, whatever the radiation and the hadronization, and the flux is a straight line in $c$ to all orders in QCD,
\begin{equation}
\Fl(\theta)=\frac{3}{4R_b}\,c\sum_h\Bn_h\,\sigma^h_A=2\,A\,c ,
\qquad
A\equiv\int_{-1}^{1}dc\;\mathrm{sgn}(c)\,\tfrac12\Fl .
\label{eq:Fexact}
\end{equation}
Its amplitude $A$ is the forward-backward asymmetry of the $b$ hadrons. It differs from the pole asymmetry $\AFBz$ by a QCD factor that depends on neither the detector nor the selection~\cite{Lee:2026bcq}, and by the standard QED and electroweak corrections, which are applied in the companion Letter.

Two further properties of the flux hold exactly, and they are the basis of the measurement. First, the bottom charge is conserved, $\int_{-1}^{1}\Fl\,dc=0$, and by Eq.~\eqref{eq:Fexact} the whole even part of the measured flux must vanish. Second, the two $b$ hadrons of an event carry opposite bottom charges at production, $\Bn_1\Bn_2=-1$; this also holds for the pair from $g\to b\bar b$. Section~\ref{sec:principle} turns these two properties into the calibration of the tagger.

We present the flux per unit polar angle rather than per unit $c$, because the detector and the tagger are binned and calibrated in $\theta$. The Jacobian $|dc/d\theta|=\sin\theta$ turns the straight line of Eq.~\eqref{eq:Fexact} into
\begin{equation}
\frac{d}{d\theta}\Big(\frac{1}{\Nbb}\sum_h\Bn_h\Big)=\sin\theta\;\Fl(\theta)=A\,\sin2\theta ,
\label{eq:sin2theta}
\end{equation}
a modulation that vanishes at $90^\circ$, has its extrema at $45^\circ$ and $135^\circ$, and integrates to zero over the full angular range. This is the shape expected in all the flux figures of this paper, and its amplitude is $A$.

\section{Measurement Principle}
\label{sec:principle}

To discuss the measurement principle, we first review the general strategy by which the LEP measurements of $\AFBz$ with the jet charge were calibrated~\cite{ALEPH:2005ab}. They calibrated two properties of their tagger on the data and took four residual corrections from the simulation. The $b$-tagging efficiency was measured in the data by counting how often both hemispheres of an event were tagged and how often only one was. This is the double-tag method developed for $R_b$, and it needs one residual correction for the correlation $C_b$ between the two hemispheres. For the jet charge, with $Q_{F/B}$ the jet charge of the forward/backward hemisphere and $Q_{b/\bar b}$ that of the quark/antiquark hemisphere, the measured forward-backward charge difference is $\langle Q_F-Q_B\rangle=\delta_b\,A^{b\bar b}_{\rm FB}$, where $\delta_b=\langle Q_b-Q_{\bar b}\rangle$ is the charge separation. It was calibrated from the product of the two hemisphere charges in the data, $\langle Q_FQ_B\rangle$, up to a hemisphere correlation $\rho_b$. The mixing of the neutral $B$ mesons reduces the charge separation and is absorbed in the measured $\delta_b$. The asymmetry was corrected for gluon radiation by $\Cqcd$. The correction is non-trivial, because the selection of a given analysis biases its sample, and the fraction that applied to each analysis was again estimated with the simulation. Finally, the contamination from lighter flavors was also estimated with the simulation.

The bottom flux follows the same principle. It keeps the two calibrations from the data and the residual corrections from the simulation. The observable itself removes $\Cqcd$, since the flux is defined at hadron level and its shape is exact, and it removes the dependence of $C_b$ and $\rho_b$ on the event topology. A measurement of the flux needs, for each $b$ hadron, its flight direction and the sign of its bottom charge at production. The double tag of the LEP measurements is recast as a tag-and-probe between the two $b$ hadrons of an event, differential in their angles. The product of the hemisphere charges becomes the fraction of same-sign pairs, whose expectation is fixed by $\Bn_1\Bn_2=-1$, and the mixing of the neutral $B$ mesons is absorbed in the counterpart of $\delta_b$, the measured dilution factor of the sign. In the following subsections we derive the methodology and discuss the measurement in detail.

\subsection{Tagging Model}
\label{sec:principle:model}

A heavy-flavor tagger returns three quantities for each hadron: whether it is found, a class $k$, and a sign $s=\pm1$. A class is a label that the tagger assigns, the same for a hadron and its antiparticle. It can be a species, such as $B^\pm$, $B^0$, or a charm hadron, or any other category the tagger distinguishes, and $\eps_k$ is the probability that a hadron is found and assigned to class $k$. For an exclusive tagger the class is the reconstructed species. For the inclusive cascade estimator used here it is the source of the sign, $L$ (the charge of a lepton in the jet of the vertex) or $Q$ (the vertex charge, $Q_{\rm vtx}\neq0$), the first available in that order. Hadrons found with neither source form the unsigned class $U$, which enters the efficiency but not the signed flux. A single signed class already gives a complete measurement.

The sign is wrong with probability $w^+_k(\theta)$ for a hadron with $\Bn=+1$ and $w^-_k(\theta)$ for a hadron with $\Bn=-1$, the mistag fractions of the class. The expectation value of the sign is then linear in $\Bn$,
\begin{equation}
E[\,s\,|\,\Bn\,]=D_k\,\Bn+\Delta_k ,\qquad
D_k\equiv1-w^+_k-w^-_k ,\qquad
\Delta_k\equiv w^-_k-w^+_k ,
\label{eq:DDelta}
\end{equation}
where $D_k$ is the dilution factor of the class, the counterpart of the charge separation $\delta_b$ at LEP and the factor by which the measured asymmetry is reduced, and $\Delta_k$ is an offset from a response of the detector that differs between positive and negative tracks. For a symmetric mistag fraction $w$, $D=1-2w$. DELPHI and SLD wrote the same factor as $2p-1$, with $p=1-w$ the probability of a correct sign~\cite{DELPHI:2004wzo,SLD:2002zsr,SLD:2004kjl}, and the flavor-tagging literature calls it the dilution~\cite{LHCb:2012dgy,Belle-II:2021zvj}. A detector response that treats both charges alike has $\Delta_k=0$. With $N^{\pm}_k(\theta)$ the number of class-$k$ hadrons with sign $\pm$ in a bin $\Delta c$, and $N_k=N^+_k+N^-_k$, the measured signed and unsigned densities per hadronic event are
\begin{equation}
\Fh_k(\theta)\equiv\frac{N^+_k-N^-_k}{\Nhad\,\Delta c},\qquad
\nh_k(\theta)\equiv\frac{N_k}{\Nhad\,\Delta c},
\label{eq:measured}
\end{equation}
where $\Nhad$ is the number of hadronic $Z$ decays in the sample. The bins are in $\theta$, and $\Delta c$ is the width of a bin in $c$, so that the measured densities are per unit $c$, like $\Fl$ and $n$. The flux per unit polar angle of Eq.~\eqref{eq:sin2theta}, shown in the figures, is $\sin\theta$ times $\Fl$.

At leading power, every species $H$ of $b$ hadron has the same angular distribution up to its production fraction $f_H$, $\Fl_H=f_H\,\Fl$ and $n_H=f_H\,n$, because the hadronization of the $b$ quark into a given species does not depend on the direction of the quark. The sum over species therefore factorizes, and
\begin{align}
E[\,\Fh_k\,]&=R_b\,\bar\eps_k(\theta)\Big[\bar D_k(\theta)\,\Fl(\theta)+\bar\Delta_k(\theta)\,n(\theta)\Big]+(\text{lighter flavors}),
\nonumber\\
E[\,\nh_k\,]&=R_b\,\bar\eps_k(\theta)\,n(\theta)+(\text{lighter flavors}),
\label{eq:expect}
\end{align}
with the class averages
\begin{equation}
\bar\eps_k\equiv\sum_Hf_H\,\eps_k(H),\qquad
\bar D_k\equiv\frac{\sum_Hf_H\,\eps_k(H)\,D_k(H)}{\bar\eps_k},
\label{eq:bars}
\end{equation}
and likewise $\bar\Delta_k$. For an exclusive tagger, the class is a species, $k=H$, and the averages are the values of that species. For an inclusive tagger, each class mixes the species, and only the averages are defined. Equation~\eqref{eq:expect} relates the true flux $\Fl$ to what is counted in each class, the raw histogram $\Fh_k$. It also shows that the measurement does not need the fractions $f_H$ of the $b$-hadron species, nor how efficiently or how reliably each of them is tagged. All of that enters only through three averages per class, $\bar\eps_k$, $\bar\eps_k\bar D_k$, and $\bar\eps_k\bar\Delta_k$, and these are exactly what the pair calibration measures.

The calibrations of both $\bar\eps_k$ and $\bar D_k$ rest on one property of pairs. For a quantity $X$ of each hadron with $E[\,X\,|\,t\,]=\lambda\,t$, and independent errors on the two hadrons of an event, $E[\,X_1X_2\,]=\lambda_1\lambda_2\,E[\,t_1t_2\,]$. For the efficiency, $X$ is ``found or not'', $t=1$ and $\lambda=\eps$. For the dilution factor, $X=s$, $t=\Bn$, and $\lambda=D$, once the offsets $\Delta$ are subtracted. Both go back to methods used at LEP and SLD, the double tag of $R_b$ for the efficiency and the same-sign fraction of double-signed events for the dilution factor. What changes is that $X$ is evaluated for each hadron at its own direction, differentially in $\theta$, and per class.

The flux does not require two tags in an event. Every tagged hadron enters the flux, and the events in which both hadrons are tagged also form the calibration sample.

\subsection{Efficiency}
\label{sec:principle:eff}

Let $N_k(\theta)$ be the number of class-$k$ hadrons at $\theta$, and $N_{kk'}(\theta,\theta')$ the number of pairs with a class-$k$ hadron at $\theta$ and its partner in class $k'$ at $\theta'$. Every tagged hadron has exactly one partner, except in the events with four $b$ hadrons, which are treated as a residual (Sec.~\ref{sec:principle:residuals}). Summing over where the partner goes gives one equation for each tag bin $\theta$,
\begin{equation}
N_k(\theta)=\sum_{\theta'}\frac{N_{kk'}(\theta,\theta')}{\bar\eps_{k'}(\theta')\,\big[1+C_{kk'}(\theta,\theta')\big]} ,
\label{eq:tnp}
\end{equation}
where $C_{kk'}$ is the correlation between the efficiencies of the two hadrons, LEP's $C_b$ for each class pair and differential in angle. The system is linear in the unknowns $1/\bar\eps_{k'}(\theta')$, one for each probe bin. The partners are mostly back-to-back, $\theta'\approx\pi-\theta$, so each equation is dominated by one cell, and each $1/\bar\eps_{k'}(\theta')$ is determined mainly by the tags in the opposite bin and their pairs. The system is therefore well conditioned. Equation~\eqref{eq:tnp} sums over the directions of the partner, which depend on the gluon radiation and on how the tag selects it. Two things are left. The first is a partner that falls outside the acceptance, where it cannot be found. The equation has no cell for it. The analysis suppresses this loss by counting the candidates over a wider range than the one in which the flux is fitted, and it estimates the remainder from the spread of the partners around the back-to-back direction, which is measured in the data (Sec.~\ref{sec:analysis:calibration}). The second is a change in the partner's efficiency, at a given $\theta'$, induced by the tag. This is the correlation $C_{kk'}$. It is a residual, predicted from the simulation (Sec.~\ref{sec:principle:residuals}). We expect it to be smaller than at LEP, where the largest part of $C_b$ came from the dependence on the event topology, which is absent when the double tag is differential in angle.

\subsection{Dilution Factor}
\label{sec:principle:mistag}

For pairs in which both hadrons are signed, let $N_{SS}$ and $N_{OS}$ be the numbers of pairs with the same and with opposite signs, for each class pair and angular cell $(\theta,\theta')$. Because $\Bn_2=-\Bn_1$, a pair has the same sign when one of the two signs is wrong, and the expected same-sign fraction is
\begin{equation}
\fSS(\theta,\theta')=\tfrac12\Big\{1-D(\theta)D(\theta')-\Cov+\big[D(\theta)\Delta(\theta')-\Delta(\theta)D(\theta')\big]\,E[\,\Bn_1\,|\,\theta,\theta'\,]\Big\} .
\label{eq:fss}
\end{equation}
It is estimated by $\hat f_{SS}=N_{SS}/(N_{SS}+N_{OS})$, and we write $\widehat{D(\theta)D(\theta')}$ for the product that Eq.~\eqref{eq:fss} gives with $\hat f_{SS}$ in place of $\fSS$. Here $D$ and $\Delta$ stand for the class averages $\bar D_k$ and $\bar\Delta_k$ of the two classes, with the indices suppressed, $\Cov$ is the covariance of the two signs being right or wrong, each counted as $\pm1$, for given true bottom charge, and $E[\,\Bn_1\,|\,\theta,\theta'\,]$ is the mean bottom charge of the hadron at $\theta$ in pairs whose other hadron is at $\theta'$, the local asymmetry. The last term is first order both in the charge offsets $\Delta$ and in the local asymmetry, $|E[\,\Bn\,|\,\theta\,]|=|\Fl/n|\leq\tfrac43\,A\approx0.13$. It is at the per-mille level relative to $D(\theta)D(\theta')$ and is applied as a correction with $\Delta$ known. Detector mistag and, for the neutral mesons signed by a lepton, oscillation both make a pair same-sign. The fraction $\hat f_{SS}$ counts them together, and the product is all that the measurement needs. Over all cells, the table $\widehat{D(\theta)D(\theta')}$ is a matrix of rank one, which is fitted for $D(\theta)$ at every angle,
\begin{equation}
\widehat{D(\theta)D(\theta')}=d_\theta\,d_{\theta'},\qquad d_\theta\equiv D(\theta),
\label{eq:rankone}
\end{equation}
up to one common sign of all $d_\theta$, since $-d$ gives the same products, which is fixed by $D>0$.

The correlation $\Cov$ is the counterpart of LEP's $\rho_b$. It arises from what the two hadrons share, the event topology, which sets the energies of both, and the reconstructed primary vertex (PV). Ignoring it would bias the asymmetry by $-\Cov/2D^2$. It is a residual, estimated from the simulation and bounded by the closure test (Sec.~\ref{sec:principle:residuals}). It is expected to be smaller than $\rho_b$. The hemisphere charges of LEP were correlated by the event topology, through a gluon jet shared between the hemispheres and through the fragmentation tracks on either side of the thrust axis, whereas the two signs here are taken from the two decay vertices and their own jets.

\paragraph{Charge offsets.} Under $c\to-c$ the flux is odd (Eq.~\eqref{eq:Fexact}) and the unsigned density is even, so the measured flux splits into
\begin{equation}
\Fh_{\rm odd}=R_b\big([\bar\eps\bar D]_{\rm e}\,\Fl+[\bar\eps\bar\Delta]_{\rm o}\,n\big) ,
\qquad
\Fh_{\rm even}=R_b\big([\bar\eps\bar D]_{\rm o}\,\Fl+[\bar\eps\bar\Delta]_{\rm e}\,n\big) .
\label{eq:parity}
\end{equation}
The even offset $\bar\Delta_{\rm e}$ is measured, from the difference between $++$ and $--$ pairs, $N_{++}-N_{--}$, or from the conservation of the bottom charge, $\int\Fl=0$. The odd offset $\bar\Delta_{\rm o}$ is not. It adds to the asymmetry, and no combination of pair counts separates the two, because a pair measures the sum of the offsets of its two hadrons, which for back-to-back pairs is twice the even part. It is controlled by the calibration of the detector, here the charge-dependent term of the track momentum scale (Sec.~\ref{sec:edm4hep:momentum}).

\paragraph{Mixing.}\label{sec:principle:mixing}
A lepton sign gives the flavor at decay. For a neutral $B_q$ meson that lives a proper time $t$, the sign is flipped with probability $\tfrac12(1-\cos\Delta m_q t)$. The pairs cannot tell this flip from a wrong sign, so the dilution factor $\bar D$ that they return already includes the mixing, and dividing the flux by $\bar\eps\bar D$ corrects for both. The mixing is therefore measured in the data, as part of the dilution factor, and no external value is needed.

\subsection{Classes and Normalization}
\label{sec:principle:norm}

Summed over cells, Eq.~\eqref{eq:fss} gives one number for each class pair. With independent species on the two sides, the pair counts return exactly the class averages of Eq.~\eqref{eq:bars} that multiply the flux in Eq.~\eqref{eq:expect}:
\begin{equation}
N_{kk'}\propto\bar\eps_k\bar\eps_{k'}\,[1+C_{kk'}],\qquad
\fSS(kk')=\tfrac12\big(1-\bar D_k\bar D_{k'}-\Cov_{kk'}\big).
\label{eq:pairclass}
\end{equation}
With $m$ classes there are $m(m+1)/2$ pair types for $m$ unknowns. For $k\in\{Q,L\}$, the pairs $LL$ determine $\bar D_L$, and the pairs $QQ$ and $QL$ together determine $\bar D_Q$. With $\bar D_L$ known, $QQ$ and $QL$ are two measurements of $\bar D_Q$, and their comparison could test the mistag correlations with more data.

Each class is corrected by its own measured $\bar\eps_k\bar D_k$. The lighter-flavor term and the offset $R_b\bar\eps_k\bar\Delta_k\,n$ are subtracted from $\Fh_k$ (Eq.~\eqref{eq:expect}), and the classes are combined with inverse-variance weights and divided by $R_b$:
\begin{equation}
\Fl(\theta)=\frac{1}{R_b}\;\frac{\sum_k w_k\,\Fh_k(\theta)/\big[\bar\eps_k(\theta)\bar D_k(\theta)\big]}{\sum_k w_k},
\qquad w_k\propto\bar\eps_k\bar D_k^{2} ,
\label{eq:absolute}
\end{equation}
where the weights are the inverse variances, which for counting statistics alone are the tagging powers $\bar\eps_k\bar D_k^2$ of the classes. Each corrected class returns $R_b\Fl$, so the measurement by itself fixes the product $R_b\,A$. In principle, $R_b$ can be measured from the same double tag. In this paper, it is taken from the Particle Data Group~\cite{ParticleDataGroup:2026mpi}. By Eq.~\eqref{eq:Fexact}, $\Fl$ is a straight line in $c$. Its odd part is fitted to $2A\,c$, or equivalently to $A\sin2\theta$ per unit polar angle (Eq.~\eqref{eq:sin2theta}), over the measured angular range, and its even part must vanish.

With $N$ $b$ hadrons, the statistical precision of the asymmetry is that of $N\sum_k\bar\eps_k\bar D_k^2$ hadrons tagged perfectly, and the tagging power $\eps D^2$ is what a tagger should optimize. The calibration adds to this. The tag-and-probe uses all found hadrons, signed or not. With $N_t$ found hadrons it determines $\bar\eps$ to $\delta\eps/\eps\simeq\sqrt{(1-\eps)/(\eps N_t)}$, which propagates directly to the amplitude, while the signed hadrons among them, $N_s\le N_t$, determine the asymmetry to $\delta A/A\simeq1/(AD\sqrt{N_s})$. The calibration of the dilution factor uses only the signed pairs and enters as $D^2$. It contributes an uncertainty comparable to the counting term (Sec.~\ref{sec:analysis:uncertainties}). Calibration and measurement share the same events, so the statistical uncertainties are obtained by bootstrapping over events, and all three parts shrink with the sample size.

\subsection{Residuals}
\label{sec:principle:residuals}

The pair calibration measures the leading parts of the tagger response, the efficiency, the mistag, and the mixing, on the data. What it does not measure is listed here. Most items are the counterparts of corrections that LEP took from the simulation, and they are expected to be smaller here, because the measurement is differential in the polar angle and does not depend on the event topology. Their size for the present measurement is given in Sec.~\ref{sec:analysis:uncertainties}.
\begin{itemize}
\item \emph{Efficiency correlation} $C_{kk'}$ (Eq.~\eqref{eq:tnp}), the counterpart of LEP's $C_b$. It is predicted from the simulation. The part of $C_b$ that came from the event topology, its largest part at LEP, is absent, because the double tag is differential in the angle.
\item \emph{Mistag correlation} $\Cov$ (Eq.~\eqref{eq:fss}), the counterpart of LEP's $\rho_b$. It is taken from the simulation and bounded by the closure. The hemisphere charges of LEP were correlated by the event topology, whereas the signs here come from the decay vertices and their own jets.
\item \emph{Odd charge offset} $\bar\Delta_{\rm o}$ (Eq.~\eqref{eq:parity}), the one quantity to which the pair counts are blind, because a pair measures the sum of the offsets of its two hadrons. A response of the detector that differs between positive and negative tracks and between the forward and backward directions produces a fake odd flux, $[\bar\eps\bar\Delta]_{\rm o}\,n(\theta)$. It is a property of the detector, not of the tagger, and is controlled at track level, as at LEP and SLD, here through the charge-odd term of the momentum scale (Sec.~\ref{sec:edm4hep:momentum}). Other charge-asymmetric effects, the interactions of $K^\pm$ and photon conversions in the material, are expected to be small for $b$ hadrons, whose decay products are hard tracks.
\item \emph{Events with four $b$ hadrons} from gluon splitting. A tagged hadron then has more than one partner, and not every pair satisfies $\Bn_1\Bn_2=-1$. Their rate is small and known from the simulation.
\item \emph{Angular migration} and \emph{within-bin variation} of $\bar\eps$ and $\bar D$. Both are set by the direction resolution against the bin width and by the gradients of the efficiency and of the flux, and both are of second order.
\item \emph{Lighter-flavor background}. The lighter-flavor term of Eq.~\eqref{eq:expect} is taken from the simulation, since the pair counts constrain neither its rate nor, for charm, the sign of its decay products. Charm can instead be given classes of its own, so that it becomes a flux calibrated by the same pair counts.
\item \emph{Kinematic bias of the tagger}. Any selection on lifetime, momentum, or vertex quality changes the mix of hard and soft $b$ hadrons, which carry different local asymmetries, so that the flux of the tagged hadrons differs from the flux of all hadrons even after the correction with the efficiency and the dilution factor averaged over the bin. It can be removed by binning the measurement and the calibration in the tagger variables, so that the flux is corrected cell by cell, or estimated from the simulation.
\end{itemize}

\section{The DELPHI Detector and Data}
\label{sec:edm4hep}

In this section we describe the DELPHI detector (Sec.~\ref{sec:edm4hep:detector}), the data and simulated samples (Sec.~\ref{sec:edm4hep:samples}), their conversion to EDM4hep (Sec.~\ref{sec:edm4hep:conversion}), and the calibration of the track momentum scale derived from the converted data (Sec.~\ref{sec:edm4hep:momentum}).

\subsection{The DELPHI Detector}
\label{sec:edm4hep:detector}

The DELPHI detector and its performance are described in detail in Refs.~\cite{DELPHI:1990cdc,DELPHI:1995dsm}, and only the components used in this analysis are described here. DELPHI was a hermetic detector with a superconducting solenoid providing a uniform magnetic field of 1.23~T parallel to the beam axis throughout the volume of the tracking detectors. The tracking system consisted of a silicon Vertex Detector (VD), a jet-chamber Inner Detector (ID), and a Time Projection Chamber (TPC), which was the main tracking device. It was complemented by a drift-chamber Outer Detector (OD) in the barrel and by two sets of drift chambers, FCA and FCB, in the end caps. The VD had three layers of silicon microstrip detectors, and a particle crossed all three for $44^\circ<\theta<136^\circ$~\cite{DELPHI:1995dsm}. In 1994, the innermost and outermost layers were equipped with double-sided detectors, which measured the coordinate along the beam in addition to those in the transverse plane, and the acceptance of the innermost layer was extended from $30^\circ<\theta<150^\circ$ to $25^\circ<\theta<155^\circ$~\cite{DELPHI:1995gas,DELPHI:2003fml}.

Muons were identified with drift chambers in the barrel and the end caps, with the first layer inside the hadron calorimeter and the others outside it. Each track was extrapolated to the chambers, and its parameters were fitted to the hits associated with it~\cite{DELPHI:1995dsm}. Electrons were identified in the barrel from the energy loss in the TPC and from the shower associated with the track in the High-density Projection Chamber (HPC), using the ratio of the shower energy to the track momentum, the match in position and direction between the shower and the track, and the longitudinal shower profile. In the end caps, the ratio of the energy in the Forward Electromagnetic Calorimeter (FEMC) to the track momentum was used instead~\cite{DELPHI:1995dsm}.

\subsection{Data and Simulated Samples}
\label{sec:edm4hep:samples}

In 1994 LEP delivered the largest integrated luminosity of any year of its $Z$-pole running, all at $\sqrt{s}=91.2$~GeV~\cite{ALEPH:2005ab}. We use DELPHI's final processing of these data, called 94C2, in the short format of DELPHI's data summary tapes (DSTs). The data and DELPHI's simulation are available on the CERN Open Data portal~\cite{DELPHI:OD:short94c2,DELPHI:OD:kk2fqqpy_l94_c2}, under the collaboration's open-access policy~\cite{DELPHI:OD:policy}.
DELPHI recorded the status of each subdetector for every file of a run (``cassette'')~\cite{Glege:1996rfm}. Only runs in which the VD and the TPC were operational at more than 99\% of their nominal performance are used. The track momentum calibration of Sec.~\ref{sec:edm4hep:momentum} uses the same runs.

All simulated events were passed through DELSIM, DELPHI's detector simulation~\cite{SimulationWorkingGroup:1988sdz,SimulationWorkingGroup:1988eya}, with the description of the 1994 detector. They were then reconstructed and written to DSTs in the same way as the data, so that they also carry the processing identifier 94C2. Two sets are used.
\begin{itemize}
\item DELPHI's own simulation of hadronic $Z$ decays for 1994. The process $e^+e^-\to q\bar q$ is generated with KK2f~\cite{Jadach:1999vf,Jadach:2000ir} with initial-state radiation, and the parton shower and hadronization with PYTHIA~6.1~\cite{Sjostrand:2000wi} in DELPHI's tune~\cite{DELPHI:1996sen}. The hadron properties and decay tables are those of DELPHI's production and have not been updated since the early 2000s. In particular, every $b$ hadron has the same lifetime, 1.6~ps.
\item Samples of $Z\to b\bar b$, $c\bar c$, $s\bar s$, and $u\bar u$ plus $d\bar d$, generated with PYTHIA~8.315~\cite{Bierlich:2022pfr} in the Monash 2013 tune~\cite{Skands:2014pea}, at $\sqrt s=91.25$~GeV with initial-state radiation. The generated events are handed to DELSIM in the record format of JETSET. The $b$ and $c$ hadrons have the lifetimes and the decay tables of the generator, which follow the PDG~\cite{ParticleDataGroup:2026mpi}. The flavor fixes only the primary quarks. Gluon splitting still produces $b$ and $c$ hadrons in the lighter samples, and the truth labels are taken from the generated hadrons. The $Z\to b\bar b$ sample gives the true flux for the closure test of Sec.~\ref{sec:analysis:calibration}, and the other three describe the charm and light-quark background.
\end{itemize}

\subsection{Conversion to EDM4hep}
\label{sec:edm4hep:conversion}

In DELPHI's IDEA project, the data were reprocessed into a final short-DST format, which includes all corrections known at the time, with the aim of long-term access to the data through a C++ framework~\cite{Camporesi:2000xqf}. The framework, written in the C++ of 2000, has not kept pace with the evolution of the language and of HEP software, a role that EDM4hep now takes on. The data and simulation are converted from these short DSTs to EDM4hep with the program \texttt{delphi-edm4hep}~\cite{delphi-edm4hep:2026}. The conversion does not reprocess the data: the short DSTs are read with DELPHI's own software, and each reconstructed quantity is written unchanged to the EDM4hep type that describes it. Corrections derived in this work, such as the momentum scale of Sec.~\ref{sec:edm4hep:momentum}, are distributed separately~\cite{delphi-edm4hep:2026}. This analysis uses charged tracks, the PV, and DELPHI's standard lepton identification.

Each track is stored with its helix parameters at the point of closest approach to the beam line. The covariance matrices of the tracks and of the PV, and the lepton identification, are taken from the DSTs. The position and size of the luminous region (``beam spot'') are not stored on the DSTs. DELPHI kept them in a separate database, with an entry for every cassette. The converter reads the beam spot from this database and stores it in each event.

AABTAG, DELPHI's $b$-tagging package~\cite{Borisov:1994krw,Borisov:1997sk,DELPHI:2002tzi}, tags jets and hemispheres from $b$ quarks using the lifetime and the decay properties of $b$ hadrons. It refits the PV with the beam spot as a constraint, clusters the event into jets with the JADE algorithm at $y_{\rm cut}=0.01$~\cite{JADE:1986kta,JADE:1988xlj}, and searches each jet for secondary vertices. In hadronic $Z$ decays, a secondary vertex is reconstructed in about 44\% of the jets with a $b$ hadron, with a $b$ purity of about 85\%~\cite{DELPHI:2002tzi}. AABTAG reconstructs the secondary vertices internally, and they are not stored on the DSTs. The converter therefore runs AABTAG v7.14 on every event, and stores AABTAG's PV, the impact parameters of the tracks with respect to it, and the secondary vertices with their tracks. The $b$-hadron candidates of this analysis are selected from these secondary vertices (Sec.~\ref{sec:analysis:reco}). Objects absent from the DSTs, such as AABTAG's secondary vertices, are thus regenerated with DELPHI's preserved software.

\subsection{Charged-Track Momentum Scale}
\label{sec:edm4hep:momentum}

\begin{figure}[tbp]
\centering
\includegraphics[width=0.9\textwidth]{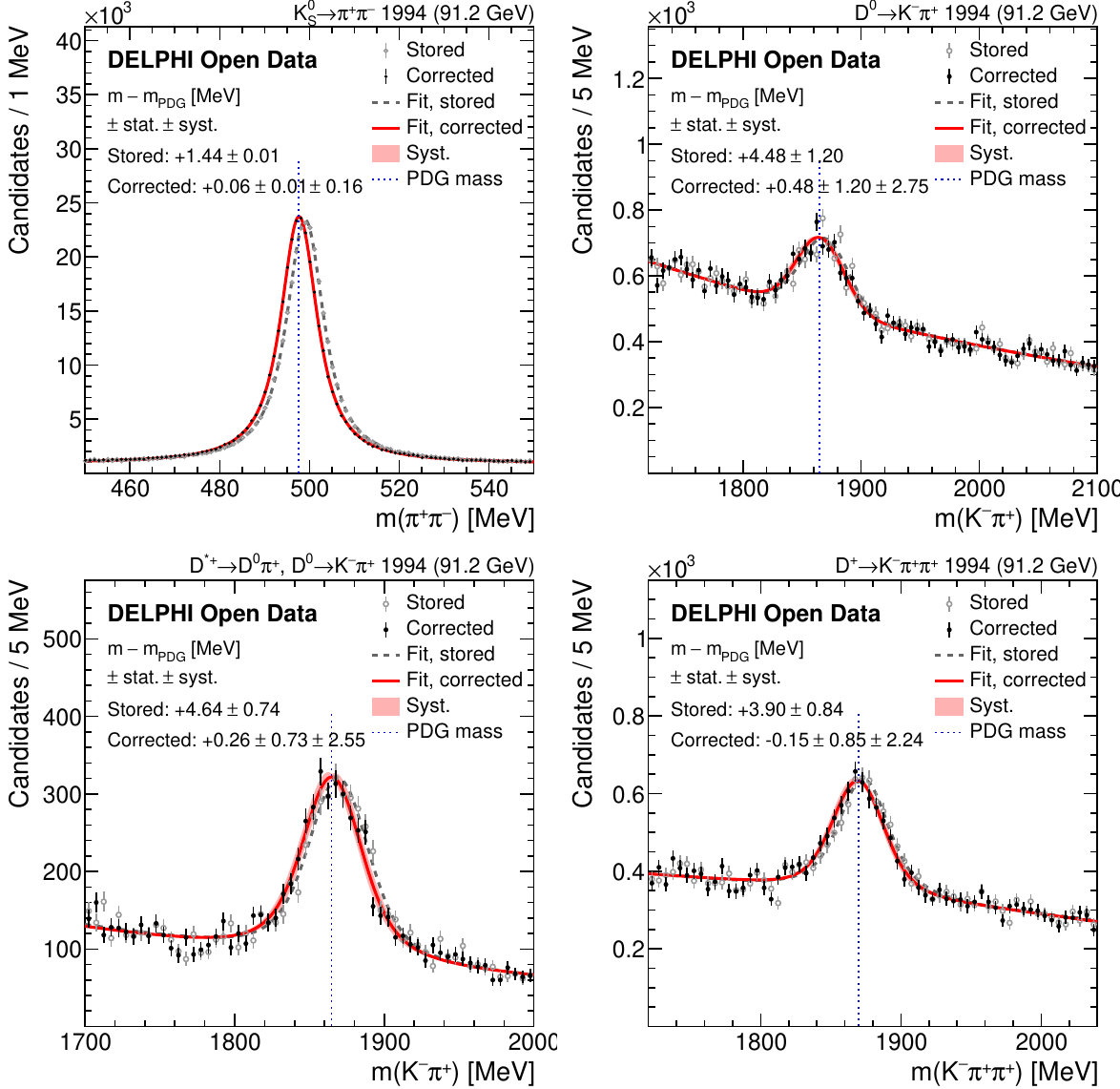}
\caption{Mass peaks of $K^0_S\to\pi^+\pi^-$, $D^0\to K^-\pi^+$ (inclusive and from $D^{*+}\to D^0\pi^+$), and $D^+\to K^-\pi^+\pi^+$ in the 1994 data, with the measured (open gray) and the corrected (filled black) momenta. The curves are fits to each distribution, and the band shows the systematic uncertainty of the correction. The vertical dotted lines mark the PDG masses, and the fitted positions relative to them are given in each panel.}
\label{fig:momscale:peaks}
\end{figure}

A charge-odd difference in measured track curvature, or a difference between the forward and backward regions, would shift the momenta of positive and negative tracks differently on each side. Such an asymmetric shift of the measured momenta can mimic a signed flux. DELPHI itself reported a shift of the measured track momenta: in a search for the $D^{*\prime}$ based on the 1992--1995 data, the $D^0$ mass was found about 4~MeV above its expected value~\cite{DELPHI:1998oyl}. A recent analysis of the DELPHI open data likewise observed a shift of the $\Xi^-$ mass in data, and corrected the reconstructed mass with a scale calibrated on the $\Lambda$ and $K^0_S$ peaks, rather than the track momenta~\cite{Fan:2026ckf}. We find the same shift in the converted 1994 data: the $D^0$ peaks lie $4.5\pm1.2$ and $4.6\pm0.7$~MeV above the PDG value, and the $K^0_S$ peak $1.44\pm0.01$~MeV above it (Fig.~\ref{fig:momscale:peaks}). In DELPHI's processing, distortions in drift time, magnetic field, and misalignment were corrected on the track elements of each subdetector, which were then refitted~\cite{Lipniacka:1995lyo}. The DSTs keep only the fitted tracks, so any remaining distortion must be corrected in the fitted momenta. We therefore calibrate the track momenta using $K^0_S\to\pi^+\pi^-$ decays and $Z\to\mu^+\mu^-$ events in data.

We correct the measured momenta with the parametrization from Ref.~\cite{CMS:2024lrd}, extended by a term for tracks from decays away from the beam line:
\begin{equation}
p_{\rm corr}=p^\prime\left(1+\kappa-\eps_{\rm loss}\,\frac{g(p/m_\pi)}{p^\prime_T}+q\,M\,p^\prime_T\right),
\qquad
p^\prime=p-\rho\,\frac{E_{\rm mat}(r_0)\,g(p/m_\pi)}{\sin\theta},
\label{eq:momscale}
\end{equation}
where $p$ is the measured momentum, $q$ the charge, and $g$ DELPHI's energy-loss shape, evaluated with the pion mass $m_\pi$. The momentum $p^\prime$ is the measured momentum after the decay-radius term is removed, and $p^\prime_T=p^\prime\sin\theta$ its transverse component. The parameters are:
\begin{itemize}
\item $\kappa$, a charge-even term, as from the modeling of the magnetic field.
\item $\eps_{\rm loss}$, a mismodeling of the energy loss in the material, which follows the energy-loss shape.
\item $M$, a charge-odd term, as from a misalignment of the tracking detectors, which shifts positive and negative tracks in opposite directions.
\item $\rho$, the normalization of a term that depends on the decay radius $r_0$. The DSTs store each track at its point of closest approach to the beam line, extrapolated inward from the first measured point, with the energy lost in the material added to its momentum. For a track from the decay at a radius $r_0$, the stored momentum thus includes the energy loss in the material inside $r_0$, which the track never crossed. DELPHI's own $V^0$ reconstruction extrapolated the tracks from their first measured point only down to $r_0$ and is not affected. The function $E_{\rm mat}(r_0)$ is the energy loss in the material inside $r_0$, taken from the material description of DELPHI's track fit.
\end{itemize}
The parameters $\kappa$ and $\eps_{\rm loss}$ are fitted separately in the backward region ($\cos\theta<-0.77$), the barrel, and the forward region ($\cos\theta>0.77$), and the parameter $M$ in 14 bins of $\theta$, $10^\circ$ wide between $30^\circ$ and $150^\circ$. A single value of $\rho$ is fitted for all tracks, and its dependence on the decay radius is given by $E_{\rm mat}(r_0)$. The $K^0_S$ candidates are divided into bins of $r_0$, $p_T$, and the region of each pion. In each bin, the mass distribution is fitted with a Voigt profile, a Gaussian convolved with a Lorentzian, on a linear background. The difference between the fitted peak position and the PDG mass~\cite{ParticleDataGroup:2026mpi} constrains the parameters.

The $Z\to\mu^+\mu^-$ events are selected by requiring two oppositely charged tracks identified as muons and an acollinearity $\theta_{\rm acol}$, the opening angle of the two tracks subtracted from $180^\circ$, below $0.15^\circ$. For each charge and $\theta$ bin, the distribution of $E_{\rm beam}/p$, with $E_{\rm beam}$ the beam energy, is fitted with a Gaussian core and power-law tails that describe the radiation, with the tails shared by the two charges.

\begin{figure}[tbp]
\centering
\includegraphics[width=0.49\textwidth]{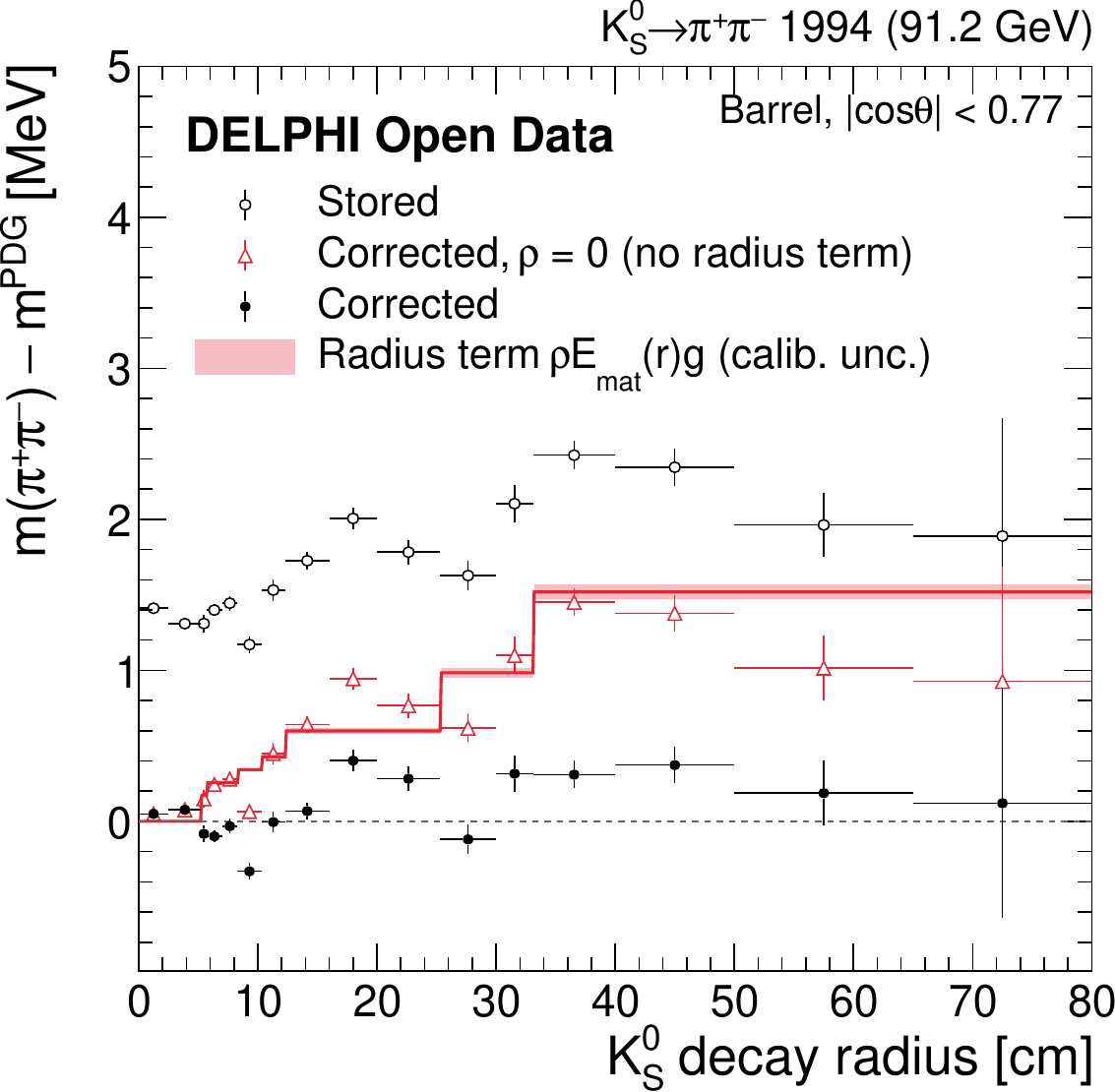}
\caption{Position of the $K^0_S$ mass peak relative to the PDG mass against the decay radius, with the momenta stored at the point of closest approach to the beam line (open circles), after the correction without the decay-radius term ($\rho=0$, open triangles), and after the full correction (filled circles). The line shows the decay-radius term of Eq.~\eqref{eq:momscale} converted to a mass shift, with its statistical uncertainty as a band.}
\label{fig:momscale:radius}
\end{figure}

The $r_0$ dependence of the $K^0_S$ sample determines $\rho$, and the wide range of pion momenta separates $\kappa$ from $\eps_{\rm loss}$. The muons at 45~GeV constrain $\kappa$ at high momentum and measure $M$ in each $\theta$ bin. The parameters are obtained in a single $\chi^2$ fit to the $K^0_S$ peak positions and the muon momenta. Figure~\ref{fig:momscale:radius} shows the effect of the $r_0$ term on the $K^0_S$ mass in the barrel. Table~\ref{tab:momscale} shows the fitted parameters.

\begin{table}[tbp]
\centering
\caption{Momentum-scale parameters of Eq.~\eqref{eq:momscale} fitted on the 1994 data, with statistical uncertainties. Negative $\kappa$ means that the measured momenta are too high. Top: $\kappa$ and $\eps_{\rm loss}$ in the three regions. Bottom: $M$ in the bins of $\theta$. The normalization of the decay-radius term is $\rho=0.671\pm0.022$, with the fit $\chi^2/\text{ndf}=1052/292$.}
\label{tab:momscale}
\begin{tabular}{lcc}
\toprule
Region & $\kappa$ ($10^{-3}$) & $\eps_{\rm loss}$ (MeV) \\
\midrule
Backward ($\cos\theta<-0.77$) & $-2.24\pm0.49$ & $-2.00\pm0.23$ \\
Barrel                        & $-2.61\pm0.06$ & $+0.79\pm0.03$ \\
Forward ($\cos\theta>0.77$)   & $-2.91\pm0.46$ & $-1.21\pm0.21$ \\
\bottomrule
\end{tabular}

\vspace{1ex}
\begin{tabular}{cc@{\hspace{2em}}cc}
\toprule
$\theta$ & $M$ ($10^{-4}$~GeV$^{-1}$) & $\theta$ & $M$ ($10^{-4}$~GeV$^{-1}$) \\
\midrule
$0^\circ$--$30^\circ$  & $-2.50\pm0.67$ & $90^\circ$--$100^\circ$  & $+0.83\pm0.13$ \\
$30^\circ$--$40^\circ$ & $-6.47\pm0.37$ & $100^\circ$--$110^\circ$ & $+1.05\pm0.13$ \\
$40^\circ$--$50^\circ$ & $-1.94\pm0.25$ & $110^\circ$--$120^\circ$ & $+0.85\pm0.14$ \\
$50^\circ$--$60^\circ$ & $-0.19\pm0.17$ & $120^\circ$--$130^\circ$ & $+0.16\pm0.17$ \\
$60^\circ$--$70^\circ$ & $+0.01\pm0.13$ & $130^\circ$--$140^\circ$ & $+1.85\pm0.27$ \\
$70^\circ$--$80^\circ$ & $+0.10\pm0.13$ & $140^\circ$--$150^\circ$ & $+4.58\pm0.37$ \\
$80^\circ$--$90^\circ$ & $-0.26\pm0.14$ & $150^\circ$--$180^\circ$ & $+2.33\pm0.74$ \\
\bottomrule
\end{tabular}
\end{table}

\begin{figure}[tbp]
\centering
\includegraphics[width=0.49\textwidth]{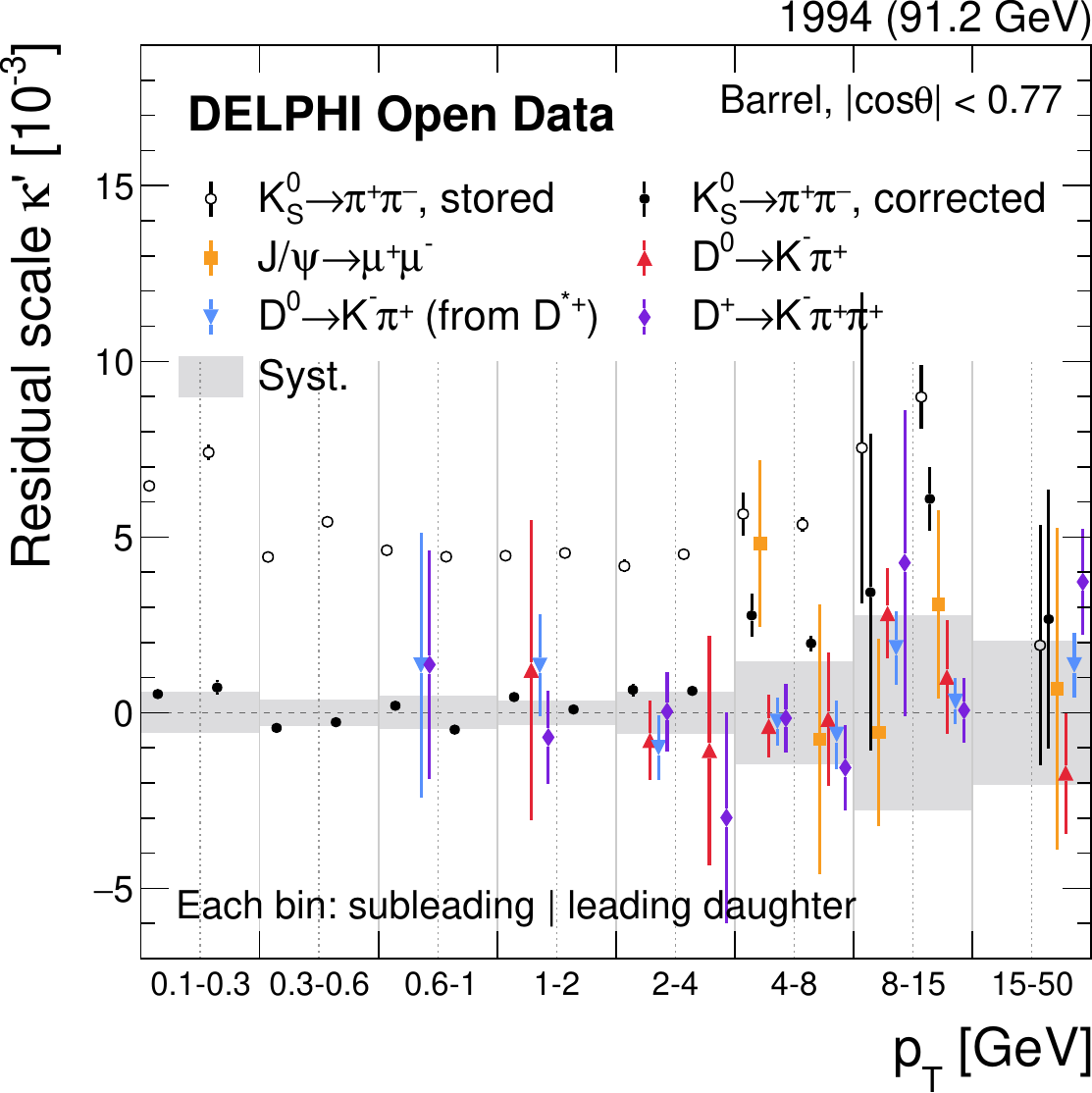}
\caption{Residual charge-even scale $\kappa^\prime$ in the barrel after the correction, for $K^0_S$, $J/\psi$, $D^0$ (inclusive and from $D^{*+}$), and $D^+$ candidates, in bins of daughter $p_T$. Each bin is split in two, with the candidates binned in the $p_T$ of the subleading (left half) or the leading (right half) daughter. The values of $\kappa^\prime$ before the correction are also shown for $K^0_S$ (open circles). The band represents the systematic uncertainty $\sqrt{\beta^2+\xi^2}$ of each bin.}
\label{fig:momscale:closure}
\end{figure}

The correction is tested on the resonances $J/\psi\to\mu^+\mu^-$, $D^0\to K^-\pi^+$ (inclusive and from $D^{*+}$ decays), and $D^+\to K^-\pi^+\pi^+$, which are not used in the calibration, and on the $K^0_S$ sample itself. A residual charge-even scale $\kappa^\prime$ and a residual charge-odd term $M^\prime$ are obtained from the peak positions after the correction. Both are measured in bins of the $p_T$ of the leading and subleading daughter. Figure~\ref{fig:momscale:closure} shows $\kappa^\prime$ in the barrel. The residuals define the systematic uncertainty of the correction. In each $p_T$ bin, the residuals of all samples are modeled as a common bias $\beta$ plus a scatter between samples $\xi$ beyond their statistical uncertainties, both fitted by maximum likelihood. The systematic uncertainty is $\sqrt{\beta^2+\xi^2}$ in the barrel. In each end cap, the uncertainty is the weighted mean of the residuals in that end cap. After the correction, the $K^0_S$ peak lies $0.06\pm0.01\,(\text{stat.})\pm0.16\,(\text{syst.})$~MeV above the PDG mass, and the $D$-meson peaks agree with their PDG masses within the statistical uncertainties (Fig.~\ref{fig:momscale:peaks}). The correction is released as a \texttt{correctionlib} file~\cite{correctionlib:software}, available through Ref.~\cite{delphi-edm4hep:2026}.

\begin{figure}[tbp]
\centering
\includegraphics[width=0.9\textwidth]{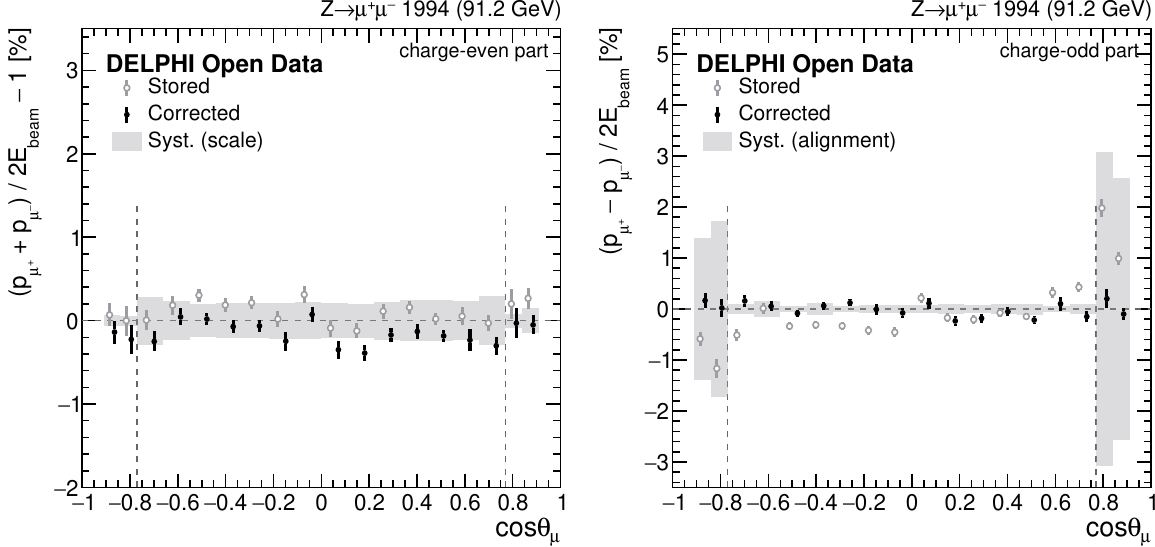}
\caption{Muon momenta in $Z\to\mu^+\mu^-$ events relative to the beam energy against $\cos\theta_\mu$, with the charge-even part $(p_{\mu^+}+p_{\mu^-})/2E_{\rm beam}-1$ (left) and the charge-odd part $(p_{\mu^+}-p_{\mu^-})/2E_{\rm beam}$ (right). Open gray points represent the measured momenta, and filled black points the corrected ones. The bands show the systematic uncertainty of the correction. The dashed vertical lines mark the boundaries between the barrel and the end caps, $|\cos\theta|=0.77$.}
\label{fig:momscale:zmumu}
\end{figure}

Figure~\ref{fig:momscale:zmumu} shows the muon momenta in $Z\to\mu^+\mu^-$ events relative to the beam energy. Before the correction, the charge-odd part reaches $+2\%$ in the forward end cap and about $-1\%$ in the backward end cap, showing an asymmetry between positive and negative tracks that differs between the two sides. After the correction, both the charge-even and the charge-odd parts are consistent with zero within the systematic uncertainty.

The converted data thus reproduce the behavior of the DELPHI data as DELPHI itself documented it, including the one deficiency found, and Appendix~\ref{app:validation} shows that they reproduce the published performance of the detector. Where the analysis needs information not stored on the DSTs, it is recovered from DELPHI's own records and software: the beam spot from DELPHI's database, the subdetector status from the cassette-level quality records~\cite{Glege:1996rfm}, and the secondary vertices from AABTAG in its final version. In the form DELPHI left them, the data support a complete physics analysis, and their conversion to EDM4hep opens them, in addition, to analysis methods and software developed since.

\section{Data Analysis}
\label{sec:analysis}

In this section we describe the analysis of the 1994 data with the method of Sec.~\ref{sec:principle}: the event selection (Sec.~\ref{sec:analysis:selection}), the reconstruction of the $b$-hadron candidates (Sec.~\ref{sec:analysis:reco}), the tagging of their bottom charge (Sec.~\ref{sec:analysis:tagging}), the in-situ calibration of the tagger (Sec.~\ref{sec:analysis:calibration}), and the statistical and systematic uncertainties (Sec.~\ref{sec:analysis:uncertainties}).

\subsection{Event Selection}
\label{sec:analysis:selection}

Hadronic $Z$ decays are selected with the standard DELPHI hadronic tag~\cite{DELPHI:2003yqh}, which was also used in DELPHI's measurements of $\AFBz$~\cite{DELPHI:2004wzo,DELPHI:2003fml} and in the earlier analyses of the DELPHI open data~\cite{Zhang:2025nlf,Zhang:2026vdo,Zhang:2026esd}. DELPHI's analysis framework SKELANA computes the tag from the charged particles of the event, and the converter stores it with each event. A charged particle enters the tag if its momentum is above 0.4~GeV, its measured track length above 30~cm, its relative momentum uncertainty below 100\%, its polar angle between $20^\circ$ and $160^\circ$, and its transverse and longitudinal impact parameters with respect to the interaction point below 4~cm and 10~cm, respectively. The tag requires at least five such particles, carrying together at least 12\% of $\sqrt s$. Neutral particles enter neither the tag nor any other step of the analysis.

We further require at least seven charged particles in the event, a summed energy of the particles accepted by the tag above 15~GeV, a center-of-mass energy within 1~GeV of 91.25~GeV, a fitted PV, and that the $b$-tagging package AABTAG (Sec.~\ref{sec:analysis:reco}) ran successfully on the event. The run selection of Sec.~\ref{sec:edm4hep:samples} removes 1.1\% of the tagged events, and the other requirements a further 1.5\%. Events with three or more $b$-hadron candidates are discarded (Sec.~\ref{sec:analysis:reco}). No thrust axis, jet, or hemisphere enters the selection. The same selection is applied to the data and to every simulated sample.

Of the 5\,753\,343 records of the 1994 data, 1\,354\,997 events pass the selection. Estimated from the simulation, the selection keeps 95.2\% of the $Z\to b\bar b$ events, 94.6\% of $Z\to c\bar c$, 94.3\% of $Z\to u\bar u$ and $d\bar d$, and 93.9\% of $Z\to s\bar s$.

\subsection{\texorpdfstring{$b$}{b}-Hadron Reconstruction}
\label{sec:analysis:reco}

The $b$-tagging tools of DELPHI already provide everything this first measurement needs. A lifetime-tagged secondary vertex gives the direction of the $b$ hadron, and its vertex charge or a lepton in its jet gives the sign. No new reconstruction is used. The one new step is to take the momentum of each vertex, rather than an event or jet axis, as the direction of the hadron.

The candidates are the secondary vertices found by AABTAG (Sec.~\ref{sec:edm4hep:conversion}). AABTAG removes the decay products of $V^0$ candidates and searches each jet for one secondary vertex among the tracks with vertex-detector hits. It starts from pairs of tracks with a common vertex of $\chi^2<4$, adds the track that increases the $\chi^2$ least as long as the increase stays below 5, and keeps the vertex if at least two of its tracks have vertex-detector hits measuring both the transverse and the longitudinal coordinates and if its distance from the PV exceeds four times its uncertainty~\cite{DELPHI:2002tzi}. Besides the tracks fitted to the vertex, AABTAG attaches tracks to it by their rapidity, and the summed momentum of the fitted and attached tracks defines the pseudo-particle momentum of the vertex.

\begin{figure}[ht!]
\centering
\includegraphics[width=0.46\textwidth]{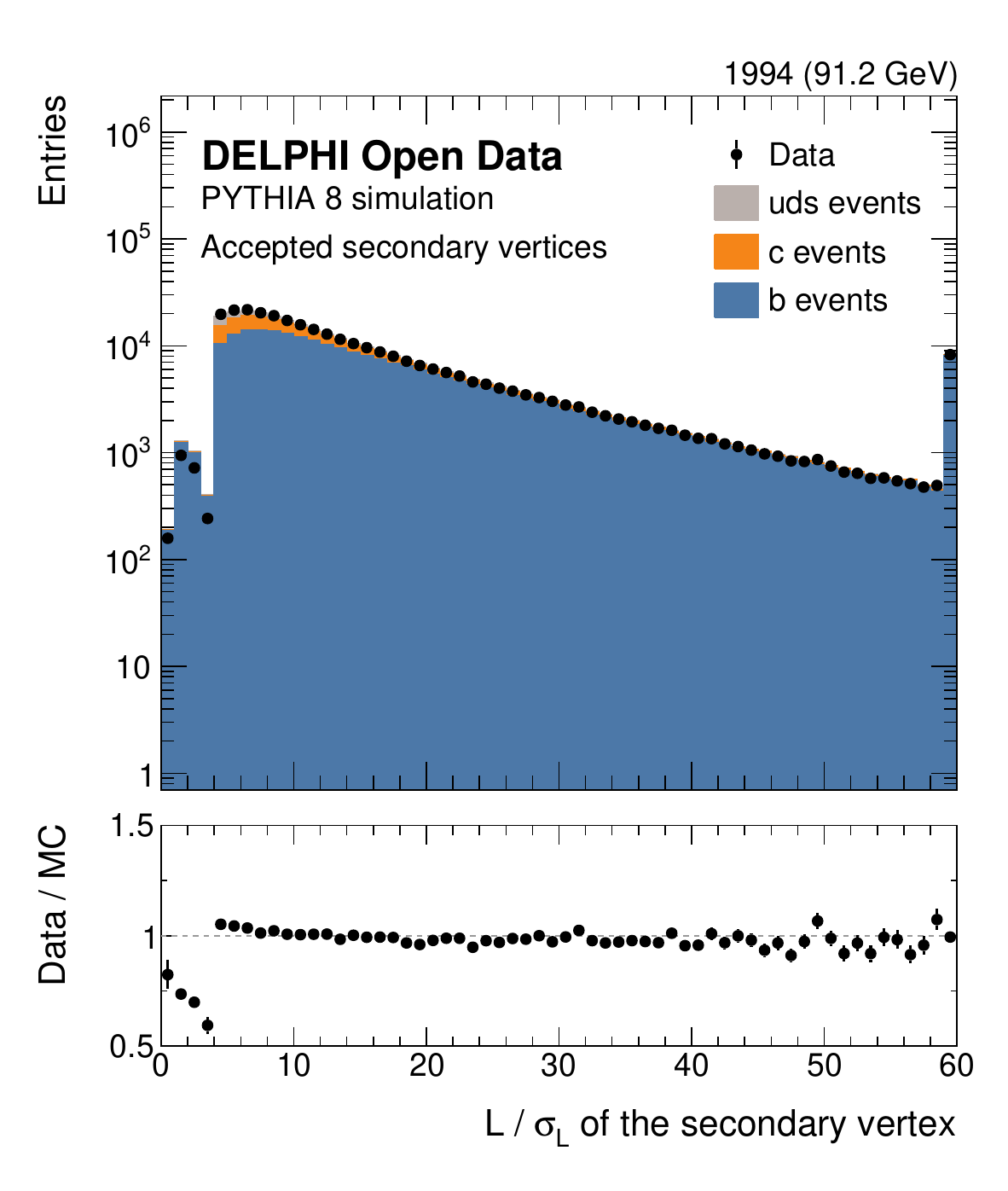}\hfill
\includegraphics[width=0.46\textwidth]{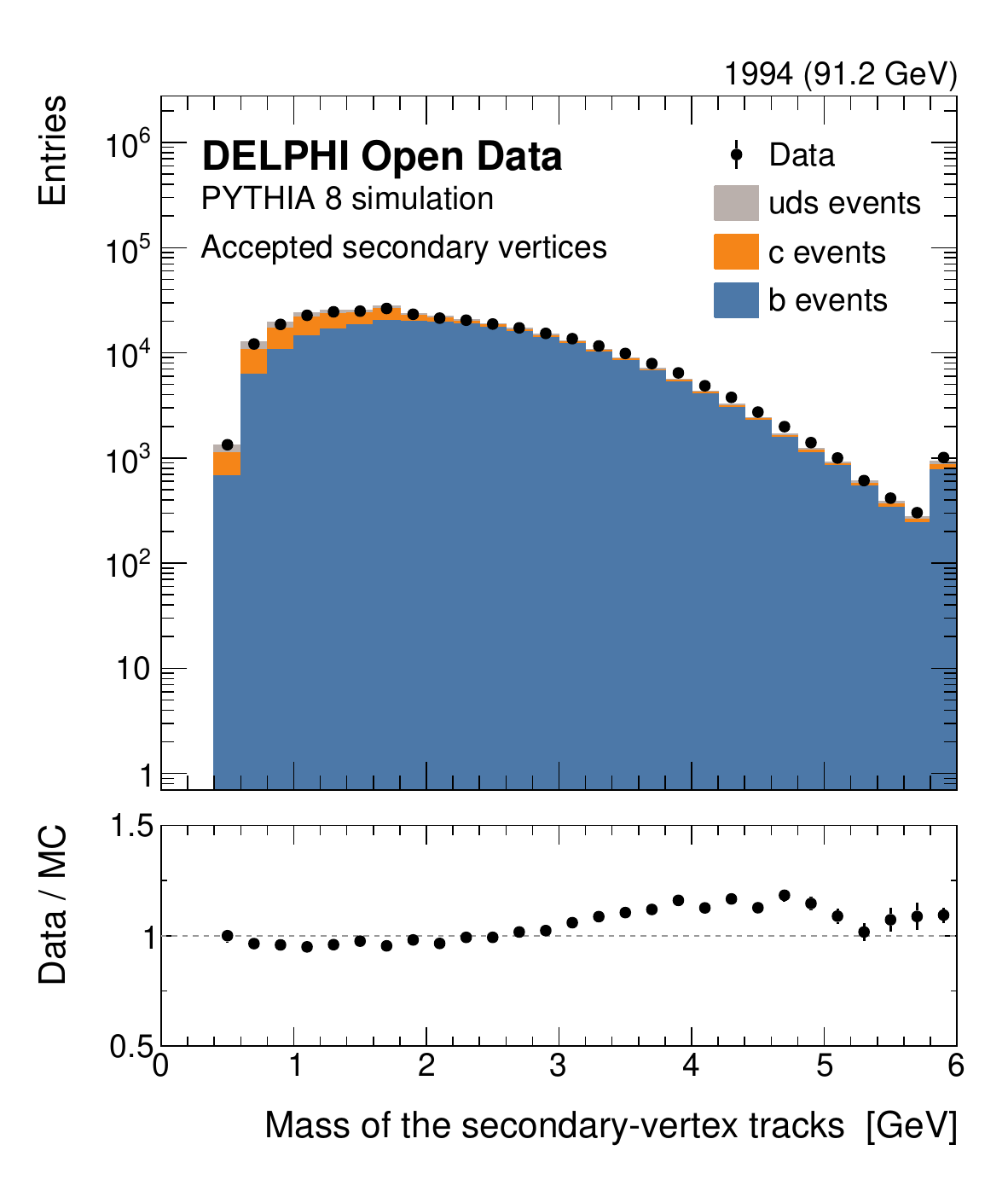}
\caption{The secondary vertices found by AABTAG in the 1994 data, before the candidate requirements, compared with the PYTHIA~8 simulation of Sec.~\ref{sec:edm4hep:samples}, stacked into $b$, $c$, and $uds$ events, with the ratio of data to simulation below. Left: significance $L/\sigma_L$ of the flight distance. Right: invariant mass $M_{\rm vtx}$ of the tracks fitted to the vertex, with the pion mass for every track. The simulation is normalized to the number of hadronic events of the data, each flavor with its partial width $R_q$.}
\label{fig:datamcvtx}
\end{figure}

\begin{figure}[ht!]
\centering
\includegraphics[width=0.46\textwidth]{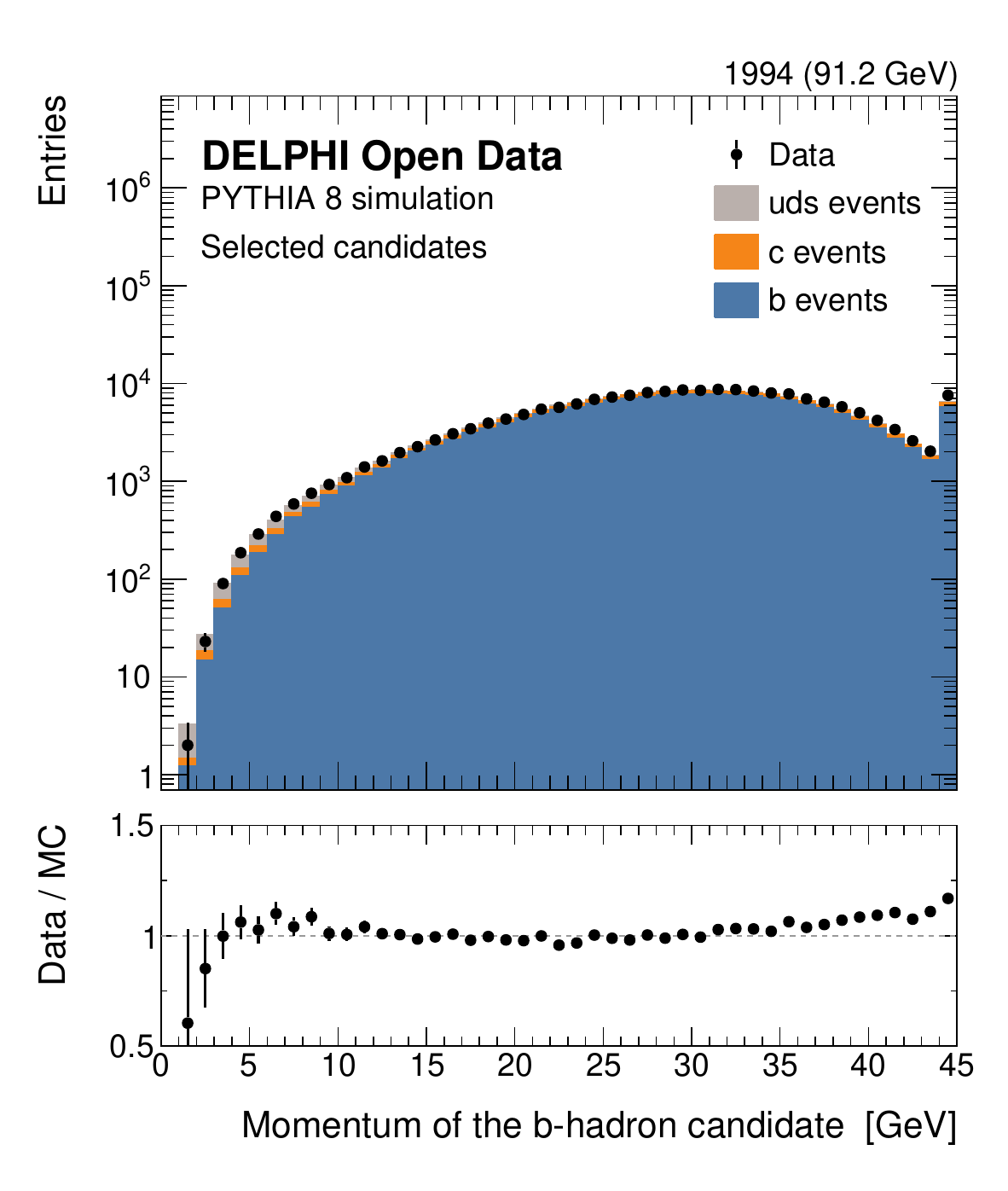}\hfill
\includegraphics[width=0.46\textwidth]{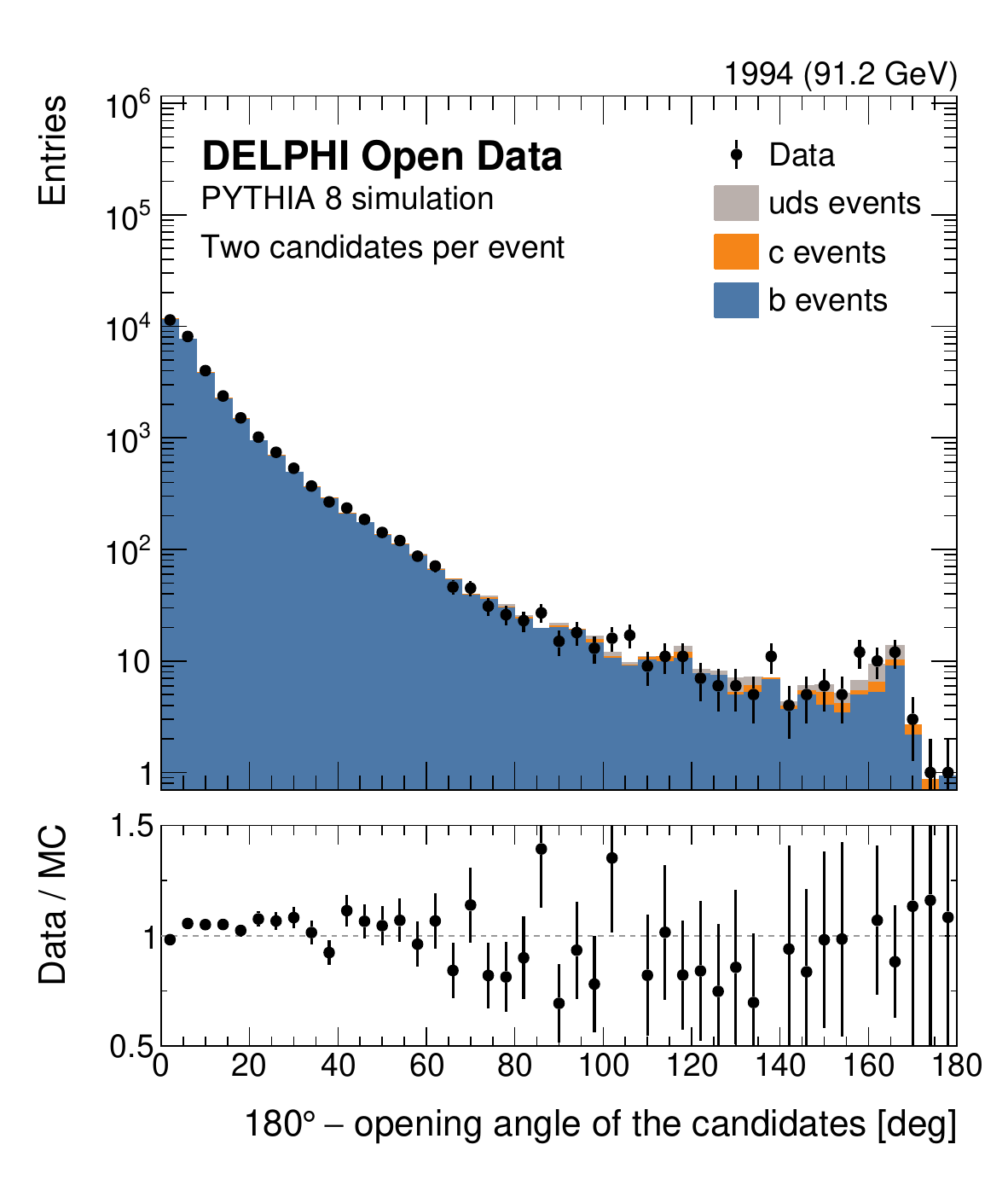}
\caption{The $b$-hadron candidates in the 1994 data compared with the simulation, as in Fig.~\ref{fig:datamcvtx}. Left: momentum of the candidates. Right: $180^\circ$ minus the angle between the two candidates, in events with two candidates.}
\label{fig:datamccand}
\end{figure}

For each vertex we compute the flight distance $L$ from AABTAG's PV and its uncertainty $\sigma_L$ from the two covariance matrices projected on the flight direction, and the invariant mass $M_{\rm vtx}$ of its fitted tracks, with the pion mass for every track. Figure~\ref{fig:datamcvtx} shows $L/\sigma_L$ and $M_{\rm vtx}$ of these vertices in the data and in the simulation. A candidate is a vertex with at least three fitted tracks, $L/\sigma_L\ge6$, and $M_{\rm vtx}\ge1.5$~GeV. The direction of the $b$ hadron is taken from the pseudo-particle momentum of the vertex, and its polar angle $\theta$ to the $e^-$ beam is the variable of the flux. In the $Z\to b\bar b$ simulation the polar angle is reconstructed without bias, with a resolution of $1.2^\circ$. Nothing else of the event enters: no thrust axis, no hemisphere, no $b$-tag of the event or of the jet, and no requirement on the number of jets or on the relative geometry of the two candidates of an event. Jets only seed AABTAG and enter the lepton selection of Sec.~\ref{sec:analysis:tagging}.

The mass requirement separates the $b$ hadrons, whose decay products reach masses of 5~GeV, from the charm hadrons and the light-quark vertices, which stay below 2~GeV. It removes 52\% of the candidates from the lighter flavors and 17\% of the $b$-hadron candidates, at little cost in statistical precision, because the low-mass vertices it removes are the ones with the smallest dilution factor. The measurement repeated without this requirement, with its own calibration, agrees with the nominal one, at the price of a larger uncertainty from the lighter-flavor templates (Sec.~\ref{sec:analysis:uncertainties}). A higher threshold would remove more charm but starve the pair calibration, whose statistical power scales with the square of the efficiency. Figure~\ref{fig:datamccand} shows the momentum of the candidates and, for events with two candidates, their acollinearity, whose tail places the partner of a candidate near $30^\circ$ or $150^\circ$ outside the range of the flux measurement (Sec.~\ref{sec:analysis:calibration}).

The 1994 data contain 191\,826 candidates in 160\,220 events: 128\,614 events with one candidate and 31\,606 with two. Events with three or more candidates, 0.03\% of the events with at least one, are discarded. In the simulation they come mostly from decay chains split into two vertices and from gluon splitting to $b\bar b$. The pairs are the input of the calibration of Sec.~\ref{sec:analysis:calibration}. In the simulation the selection finds 0.58 candidates per $Z\to b\bar b$ event, that is 29\% of the $b$ hadrons, against 0.059 per $Z\to c\bar c$ event and 0.005 per light-quark event. Weighted by the partial widths of the $Z$, these rates imply that 91\% of the candidates are $b$ hadrons, 7\% are charm hadrons, and 2\% are vertices from light-quark events. These lighter-flavor candidates are subtracted with templates from the simulated $Z\to c\bar c$, $Z\to u\bar u,d\bar d$, and $Z\to s\bar s$ samples, normalized to $R_c$, $R_u+R_d$, and $R_s$ times the number of selected events.

\subsection{Bottom-Charge Tagging}
\label{sec:analysis:tagging}

The tagger has to estimate the sign of the bottom charge of each candidate at production. Two decay products carry it here (Table~\ref{tab:sources}). A charged $B$ meson carries it as its own charge, which the vertex charge measures. The lepton of a semileptonic decay carries it through $b\to c\,\ell^-\bar\nu$. A $b$ quark gives a negative charge in both cases, so the estimated bottom charge is minus the measured charge. Each estimator is wrong with some probability: a lepton from the cascade $b\to c\to\ell^+$ has the wrong sign, and for neutral mesons mixing flips the lepton sign with the time-integrated probability $\chi$. Rather than corrected one by one, they are all absorbed in the dilution factor $D_k$ of each class, which the pair calibration of Sec.~\ref{sec:analysis:calibration} measures. The kaon of the $b\to c\to s$ chain also carries the sign, through $K^-$, but it is not used in this measurement: it needs the particle identification of the RICH detectors, whose calibration on the converted data is left for future work.

\begin{table}[ht!]
\centering
\caption{Decay sources of the two sign estimators, by $b$-hadron species. Bold: the channels that carry the bottom charge at production in each class of the cascade estimator ($Q$: the hadron's own charge, $B^\pm$ only; $L$: the lepton of the semileptonic decay). Inclusive rates per $b$ hadron: semileptonic $\approx10.7\%$ per lepton flavor, cascade $\approx8.0\%$. Mixing flips the sign of $L$ for neutral mesons with time-integrated probability $\chi$ and is inside $\bar D_k$, never corrected. The right-sign fraction of each class is not computed from these rates. It is what the pair calibration of Sec.~\ref{sec:principle:mistag} measures.}
\label{tab:sources}
\footnotesize
\setlength{\tabcolsep}{3pt}
\newcolumntype{P}[1]{>{\raggedright\arraybackslash}p{#1}}
\begin{tabular}{@{}P{0.10\textwidth}P{0.03\textwidth}P{0.04\textwidth}P{0.045\textwidth}P{0.19\textwidth}P{0.14\textwidth}P{0.09\textwidth}@{}}
\toprule
Hadron & $\Bn$ & $f_H$ & $Q_{\rm vtx}$ ($Q$) & Semileptonic ($L$, right sign) & Cascade ($L$, wrong sign) & Mixing \\
\midrule
$B^-\;(b\bar u)$ & $+1$ & $0.41$ & $\boldsymbol{-1}$ & $\boldsymbol{B^-\to D^0\ell^-\bar\nu X}$ & $D^0\to K^-\ell^+\nu$ & none \\
$\bar B^0\;(b\bar d)$ & $+1$ & $0.41$ & $0$ & $\boldsymbol{\bar B^0\to D^{(*)+}\ell^-\bar\nu}$ & $D^+\to\bar K^0\ell^+\nu$ & $\chi_d{=}0.19$ \\
$\bar B^0_s\;(b\bar s)$ & $+1$ & $0.10$ & $0$ & $\boldsymbol{\bar B^0_s\to D_s^+\ell^-\bar\nu}$ & $D_s^+\to\phi\,\ell^+\nu$ & $\chi_s{=}0.50$ \\
$\Lambda_b\;(udb)$ & $+1$ & $0.09$ & $0$ & $\boldsymbol{\Lambda_b\to\Lambda_c^+\ell^-\bar\nu}$ & $\Lambda_c^+\to\Lambda\,\ell^+\nu$ & none \\
\midrule
$B^+\;(\bar b u)$ & $-1$ & $0.41$ & $\boldsymbol{+1}$ & $\boldsymbol{B^+\to\bar D^0\ell^+\nu X}$ & $\bar D^0\to K^+\ell^-\bar\nu$ & none \\
$B^0\;(\bar b d)$ & $-1$ & $0.41$ & $0$ & $\boldsymbol{B^0\to D^{(*)-}\ell^+\nu}$ & $D^-\to K^0\ell^-\bar\nu$ & $\chi_d{=}0.19$ \\
$B^0_s\;(\bar b s)$ & $-1$ & $0.10$ & $0$ & $\boldsymbol{B^0_s\to D_s^-\ell^+\nu}$ & $D_s^-\to\phi\,\ell^-\bar\nu$ & $\chi_s{=}0.50$ \\
$\bar\Lambda_b\;(\bar u\bar d\bar b)$ & $-1$ & $0.09$ & $0$ & $\boldsymbol{\bar\Lambda_b\to\bar\Lambda_c^-\ell^+\nu}$ & $\bar\Lambda_c^-\to\bar\Lambda\,\ell^-\bar\nu$ & none \\
\bottomrule
\end{tabular}
\end{table}

Here we use two estimators, tried in a fixed order. The first one available gives the candidate its sign and defines its class $k$. Each of them is built from the candidate's own vertex and jet, never from the rest of the event, so that the pair calibration of Sec.~\ref{sec:principle:mistag} applies class by class.
\begin{itemize}
\item \emph{Lepton} ($L$): a track in the jet of the vertex with the standard muon or electron tag~\cite{DELPHI:1995dsm}, with a transverse momentum above 1~GeV with respect to the jet axis, computed after removing the lepton from the jet. No other requirement is applied. The transverse-momentum requirement keeps the soft leptons of cascade decays, which carry the wrong sign, from taking precedence over the vertex charge. The sign is minus the lepton charge, following the semileptonic decay $b\to\ell^-$.
\item \emph{Vertex charge} ($Q$): the total charge of the tracks fitted to the vertex, when it is different from zero. For a charged $b$ hadron it has the sign of the charge of the $b$ quark. The sign is minus the sign of the sum.
\end{itemize}
The lepton is tried first because it is the more reliable of the two: in the simulation its sign agrees with the flavor of the hadron at its decay for 90\% of the lepton-tagged candidates, against 70\% for the vertex charge. Candidates with neither tag form the unsigned class $U$ of Sec.~\ref{sec:principle:model}. They count in the candidate density and in the efficiency, but not in the signed flux.

In the 1994 data, 10\% of the candidates are signed by a lepton and 61\% by the vertex charge, and the remaining 29\% are unsigned. The classes also differ in their background: 2\% of the candidates in $L$, 11\% in $Q$, and 8\% of the unsigned ones come from the lighter flavors. This is why the background subtraction and the calibration are done class by class. With the lighter flavors subtracted, and the number of $b$ hadrons taken as $2R_b$ times the number of selected events, the data contain 0.30 candidates per $b$ hadron, against 0.29 in the simulation. Of the $b$ hadrons, 3\% are signed by a lepton, 18\% by the vertex charge, and 9\% are found but unsigned.

As a check of the selection, the lepton threshold was moved from 1~GeV to 0.5 and to 2~GeV, which changes the lepton class from 10\% to 18\% and to 4\% of the candidates. Each variation is calibrated from its own pair counts and bootstrapped like the nominal selection (Secs.~\ref{sec:analysis:calibration} and \ref{sec:analysis:uncertainties}). The amplitudes agree with each other within their uncertainties.

\begin{figure}[ht!]
\centering
\includegraphics[width=0.5\textwidth]{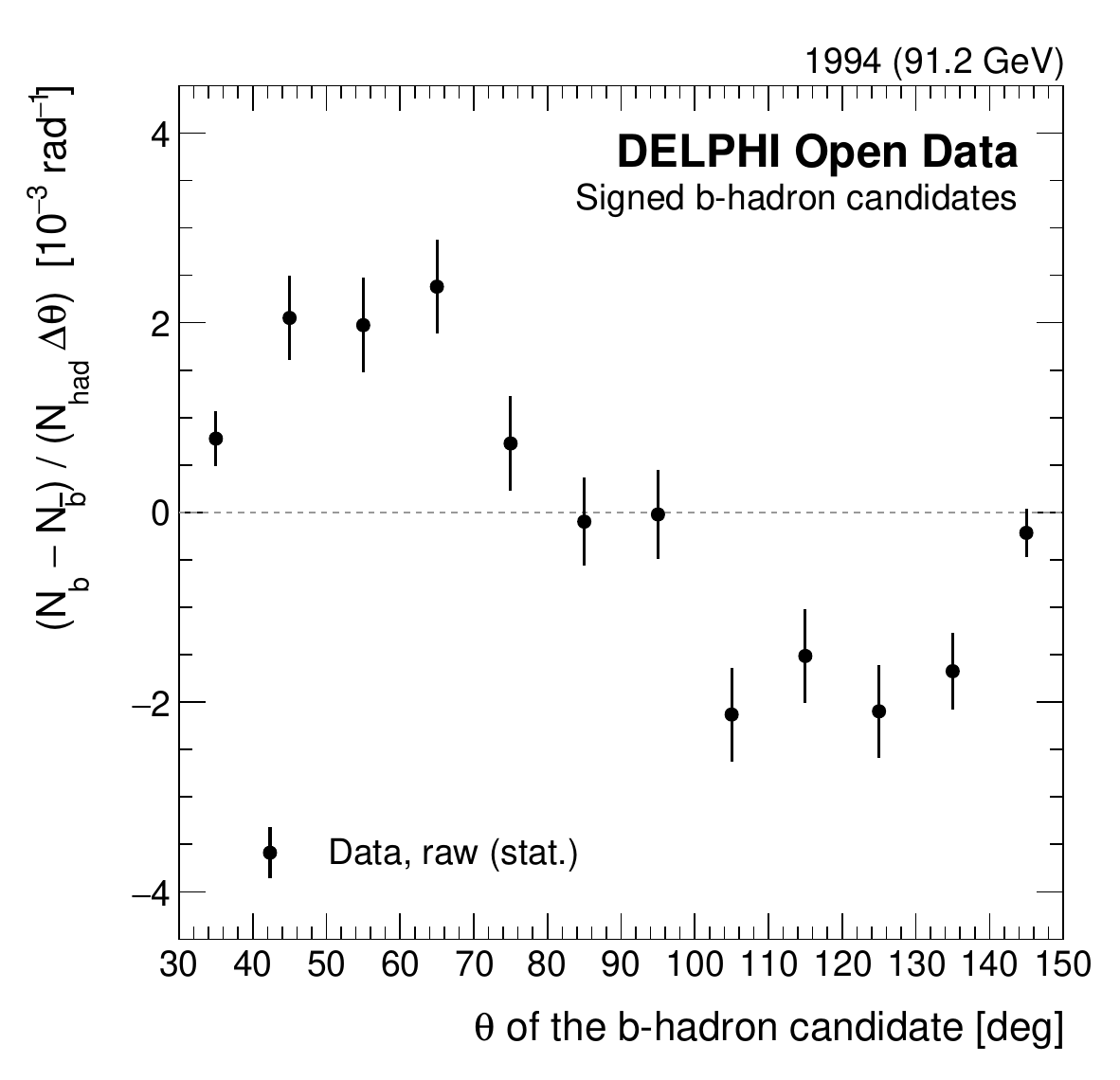}
\caption{Raw signed count of the $b$-hadron candidates in the 1994 data, $(N_b-N_{\bar b})$ per hadronic event and per radian of $\theta$, before any calibration ($\theta=0$ along the $e^-$ beam; statistical uncertainties). $N_b$ ($N_{\bar b}$) counts the candidates signed as containing a $b$ ($\bar b$) quark.}
\label{fig:rawdata}
\end{figure}

Figure~\ref{fig:rawdata} shows the result before any calibration: the signed count $N_b-N_{\bar b}$ of the candidates per hadronic event and per radian of $\theta$, with $N_b$ ($N_{\bar b}$) the candidates signed as containing a $b$ ($\bar b$) quark. The modulation in $\sin2\theta$ of Eq.~\eqref{eq:sin2theta} is already visible. Its amplitude is diluted by the efficiency and the dilution factor of the tagger, which the next section measures.

\subsection{In-Situ Calibration}
\label{sec:analysis:calibration}

The calibration follows Sec.~\ref{sec:principle} step by step. The flux is measured and fitted in twelve bins of $\theta$ between $30^\circ$ and $150^\circ$, the range that the vertex detector covers and where the vertex reconstruction performs best. The candidates, however, are accepted over the standard track acceptance of DELPHI, $20^\circ$ to $160^\circ$ (Sec.~\ref{sec:analysis:selection}), and the single and pair counts are binned in $\theta$ over this whole range. The sum over the angle of the partner in Eq.~\eqref{eq:tnp} then reaches the partners of the candidates near the edges of the measurement, a part of which gluon radiation places below $30^\circ$ or beyond $150^\circ$. The efficiencies in the outer bins are poorly determined. They enter only through the few pairs that reach these bins, and their uncertainty is propagated into the edge bins of the flux. In practice, Eq.~\eqref{eq:tnp} is not solved bin by bin, because a partner outside the acceptance has no cell and the cells at the edges are sparse. We use the same relation summed over the bins of the partner: for each tag bin, the fraction of tags whose partner is found is the efficiency averaged over the directions of the partner. The weights of this average are the spread of the partner around the back-to-back direction, measured in the data from tags between $60^\circ$ and $120^\circ$ and taken to be the same at all polar angles, times the measured angular density of the hadrons, set to the average of the measured bins where no candidate is found. They are normalized over all directions, so that the partners outside the acceptance are included. The efficiencies $\bar\eps_k(\theta)$ follow from a weighted least-squares fit.

\begin{figure}[ht!]
\centering
\includegraphics[width=0.45\textwidth]{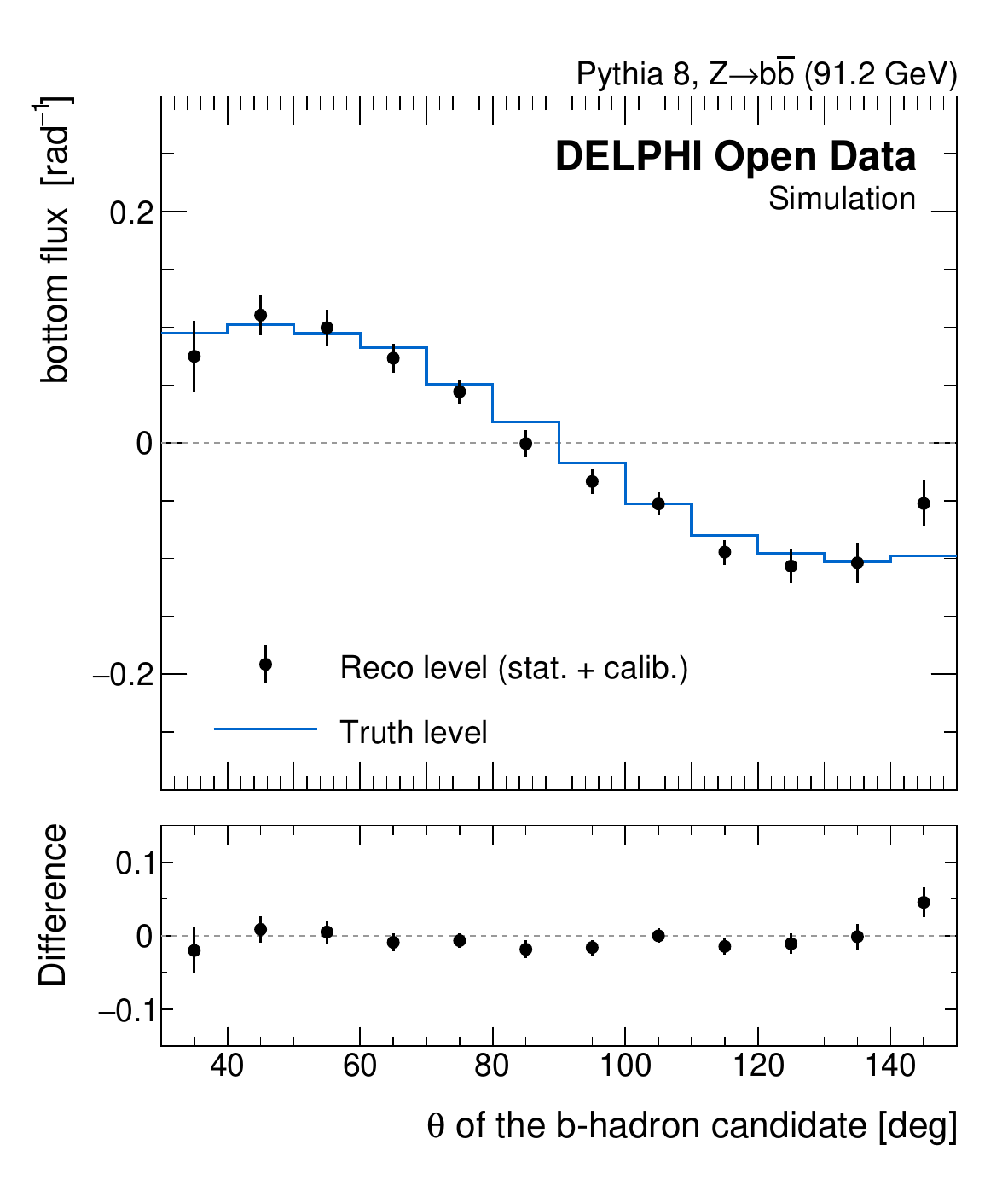}
\caption{Closure on the PYTHIA~8 $Z\to b\bar b$ simulation of Sec.~\ref{sec:edm4hep:samples}: the flux obtained at reconstruction level with the same chain as for the data, including the calibration from the pair counts (points; statistical and calibration uncertainties), and the true flux of the $b$ hadrons at production (line), with their difference below. The fitted amplitude is $1.021\pm0.048$ times the true one.}
\label{fig:closure}
\end{figure}

The dilution factor $\bar D_k(\theta)$ comes from the same-sign fraction of the signed pairs, Eq.~\eqref{eq:fss}, through the rank-one fit of Eq.~\eqref{eq:rankone}. The dilution factor of the vertex-charge class is fitted jointly to the $QQ$ and $QL$ pairs, and that of the lepton class to the $LL$ pairs (Eq.~\eqref{eq:pairclass}). The even charge offset $\bar\Delta_{\rm e}$ is measured from the difference between the $++$ and $--$ pairs. It is $0.006\pm0.006$ for $Q$ and $-0.011\pm0.020$ for $L$, both consistent with zero. The efficiency correlation $C_{kk'}$ of Eq.~\eqref{eq:tnp} and the mistag correlation $\Cov$ of Eq.~\eqref{eq:fss} are set to zero. Both are residuals, evaluated in Sec.~\ref{sec:analysis:uncertainties}, and the closure test on the simulation (Fig.~\ref{fig:closure}) bounds their combined effect.

No requirement is made on the opening angle of a pair. The lighter-flavor templates of Sec.~\ref{sec:analysis:reco} and the offset term are subtracted from each class, each class is divided by its own $\bar\eps_k\bar D_k$, and the classes are combined bin by bin with inverse-variance weights, Eq.~\eqref{eq:absolute}, where the variance of a class includes the uncertainty of its calibration. Before the combination, the flux of each class is checked against the others: the two agree with each other and with the combination within their uncertainties, as Eq.~\eqref{eq:expect} requires, since every class measures the same flux.

\begin{figure}[ht!]
\centering
\includegraphics[width=0.48\textwidth]{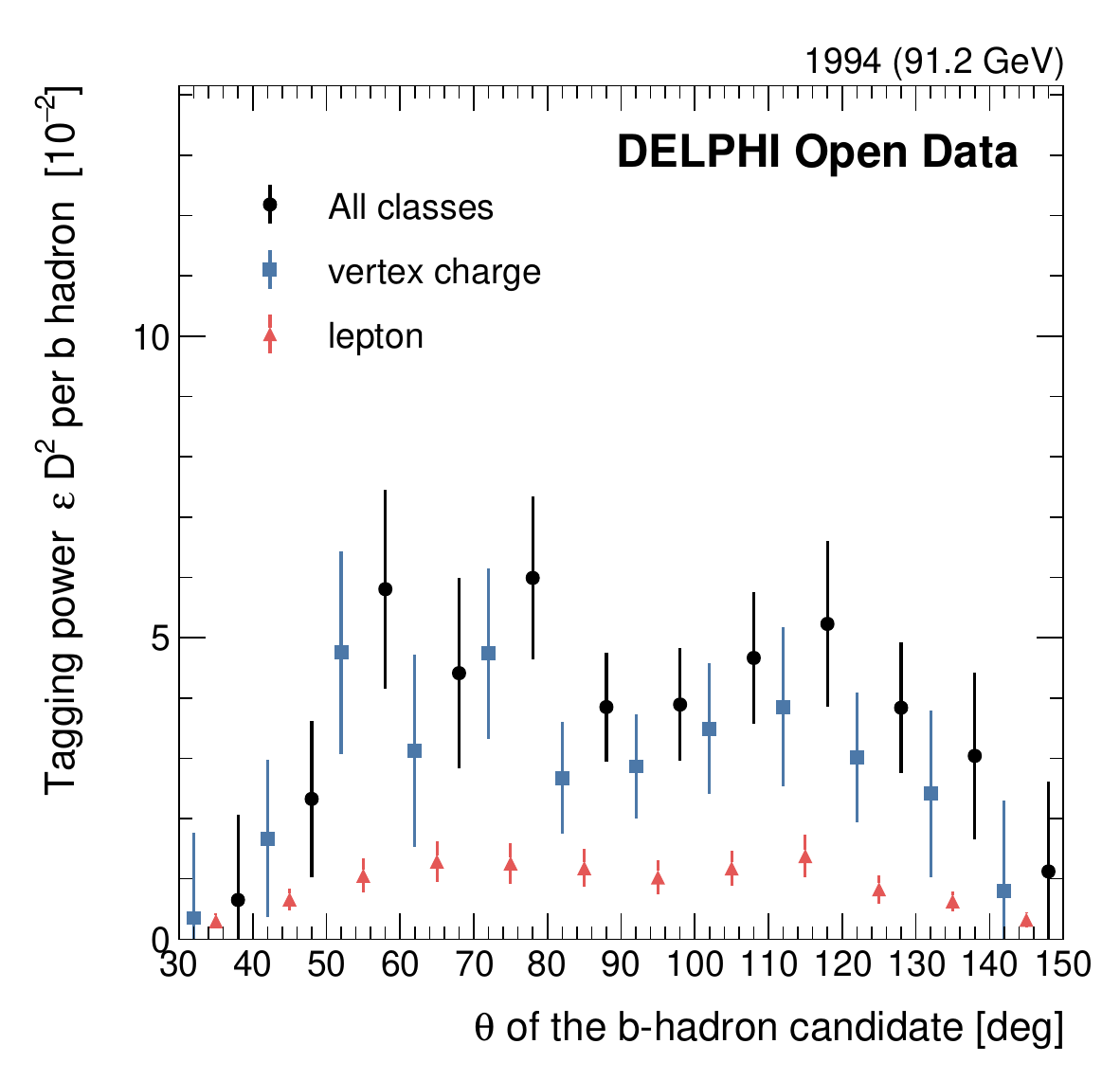}\hfill
\includegraphics[width=0.48\textwidth]{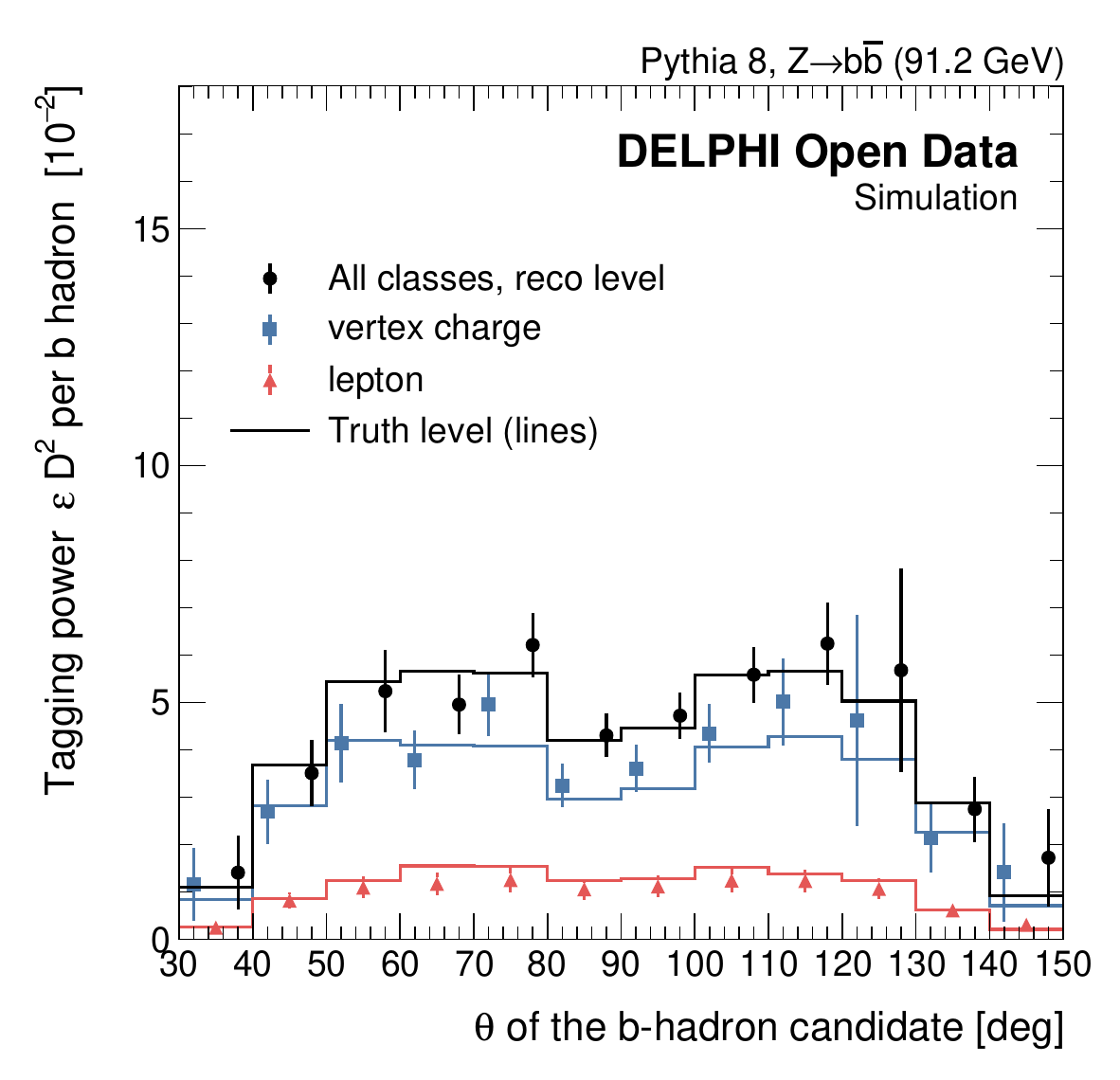}
\caption{Tagging power $\bar\eps_k\bar D_k^2$ per $b$ hadron and per bin of $\theta$, for each class and for all classes combined, obtained from the pair counts. Left: the 1994 data. Right: the PYTHIA~8 $Z\to b\bar b$ simulation, from the pair counts at reconstruction level (points) and from the truth (lines).}
\label{fig:tagpower}
\end{figure}

Figure~\ref{fig:tagpower} shows the tagging power of each class, measured in the data and simulation. The vertex charge carries most of it. It signs 61\% of the candidates, but with a small dilution factor: it is informative only for the charged $b$ hadrons, and a lost or wrongly attached track changes the charge of a vertex by one unit. For the lepton it is the other way round. Its dilution factor is large, because the semileptonic decay is clean, but its efficiency is small, set by the semileptonic branching fraction, the lepton identification, and the transverse-momentum requirement. Combined, the two classes reach a tagging power of about $5\times10^{-2}$ per $b$ hadron in the central region, falling toward $30^\circ$ and $150^\circ$ with the acceptance of the vertex detector. In the simulation, the tagging power obtained in this way, from reconstructed quantities alone, agrees with the one computed from the true flavor of each hadron, for every class and every bin. This is the closure of the calibration itself. In the data, the power per $b$ hadron is 15\% lower than in the simulation. The difference is in the vertex-charge class, whose dilution factor is $0.366$ in the data against $0.418$ in the simulation, a difference of about two standard deviations. The lepton class is, if anything, more powerful in the data ($0.500$ against $0.478$, with mixing inside both). A smaller dilution factor of the vertex charge alone points to the track content of the vertices, that is, the tracking efficiency, fake tracks, and fragmentation tracks attached to the vertex, or to the species and lifetime mix of the simulation. This does not affect the measurement, which uses the calibration from the data. It only sets its statistical precision. The power is limited by the estimators themselves, which give each candidate one source and a hard $\pm1$ sign. A dedicated tagger that combines all the tracks of the vertex, with their identification and kinematics, into one probability per candidate, and enters the flux with a weight instead of a sign, raises the power considerably (Sec.~\ref{sec:summary}).

Figure~\ref{fig:closure} shows the closure of the whole chain on the $Z\to b\bar b$ simulation: the flux obtained at reconstruction level, with the selection, the tagging, and the calibration applied exactly as to the data and nothing taken from the truth, compared with the true flux of all weakly decaying $b$ hadrons at production. A fit of the shape of Eq.~\eqref{eq:sin2theta} to the twelve bins, with the full covariance of the calibrated points, returns $A=0.1048\pm0.0049$ for the calibrated flux against $0.1027$ for the true flux, a ratio of $1.021\pm0.048$, where the uncertainty combines the statistics of the sample and of the calibration. The bin-by-bin differences are consistent with their uncertainties, $\chi^2=14.8$ for 12 bins. They are largest at the edges, $-0.020\pm0.031$ at $35^\circ$ and $+0.045\pm0.020$ at $145^\circ$, where the pair table is sparse. The partner of a candidate at one edge is usually at the opposite edge, where the efficiency is lowest. An edge cell therefore holds about 900 pairs, against 7\,000 in the center, and the dilution factor of the vertex-charge class is known to $\pm0.12$ at the edges, against $\pm0.04$ inside. The ratio is consistent with one, and the residuals of Sec.~\ref{sec:analysis:uncertainties} are too small for this closure to resolve.

\subsection{Statistical and Systematic Uncertainties}
\label{sec:analysis:uncertainties}

\paragraph{Statistical uncertainty.} The measurement and the calibration use the same events, the former through every candidate and the latter through the pairs, so their statistical fluctuations are correlated. The statistical uncertainties are therefore obtained by a bootstrap: the whole chain, from the pair counts to the corrected flux, is repeated on 2000 resamples of the selected events, and the spread of the results is the statistical uncertainty of every quantity in this paper. It contains the counting fluctuations of the signed candidates and the fluctuations of $\bar\eps_k$, $\bar D_k$, and $\bar\Delta_{\rm e}$. On the amplitude of the flux, the counting fluctuations give 9.0\% and the calibration 5.5\% (Table~\ref{tab:syst}). The parts are correlated through the shared events. The full resampling gives 10.5\% for the two combined, compatible with their sum in quadrature, so the correlation is negligible at the present precision. The calibration part is dominated by the dilution factor of the vertex-charge class, which is measured from the same-sign fractions of the $QQ$ and $QL$ pairs with $D_Q^2\approx0.1$. It shrinks with the sample size like the counting part, since both come from the data.

\paragraph{Residuals.} Of the residuals of Sec.~\ref{sec:principle:residuals}, the odd charge offset $\bar\Delta_{\rm o}$ is the one that the closure does not bound, because the simulation does not contain it. The charge-odd term of the momentum scale is measured on the converted data and corrected (Sec.~\ref{sec:edm4hep:momentum}), with a residual consistent with zero, and the decay products of $b$ hadrons are hard tracks, for which the other charge asymmetries of the detector are far smaller than for the soft tracks of the electric-charge flux~\cite{Zhang:2026vdo}. The other residuals are estimated from the simulation.

The candidates from the lighter flavors, 7\% from charm and 2\% from light-quark events (Sec.~\ref{sec:analysis:reco}), are subtracted with templates from the simulation. The charm template is a signed flux and not only a rate, since charm hadrons carry an asymmetry of their own and the same decay products that sign a $b$ hadron, $c\to\ell^+$, and the charge of a $D^\pm$ vertex sign a charm hadron with the opposite correlation. Replacing the PYTHIA~8 templates with those of DELPHI's KK2f simulation (Sec.~\ref{sec:edm4hep:samples}), whose lifetimes and decay tables differ, changes the amplitude by $(-1.6\pm 2.8)\%$, where the uncertainty is that of the statistics of the simulated samples. The change is compatible with zero and is taken in full as the systematic uncertainty. For a future measurement with a tagger that also labels charm, the charm template can be constrained in situ like the $b$ classes.

The efficiency correlation $C_{kk'}$, predicted from the truth of the simulated $Z\to b\bar b$ samples, would change the amplitude by $(-2.4\pm0.5)\%$ with PYTHIA~8 and by $(-0.6\pm1.2)\%$ with the KK2f sample, with which the closure of Sec.~\ref{sec:analysis:calibration} is consistent. The mistag probabilities of the two candidates of a pair may be correlated within an event. Measured from the right and wrong signs of every candidate pair of the simulation relative to the product of the dilutions, $\Cov/(\bar D\bar D')$ is $+0.05\pm0.03$ for the vertex-charge pairs and $-0.04\pm0.03$ for the vertex-charge--lepton pairs with PYTHIA~8, and $-0.08\pm0.05$ and $-0.12\pm0.05$ with the KK2f sample. With PYTHIA~8 the two cancel and the amplitude is unchanged. With the KK2f sample, which has three times fewer $b$ events, both are negative and add to a shift of $(+5.7\pm2.2)\%$. Table~\ref{tab:syst} quotes the PYTHIA~8 values, from the larger sample with current $b$-hadron lifetimes and decay tables (Sec.~\ref{sec:edm4hep:samples}). The two samples differ by 2.3 standard deviations. The closure test is consistent with either value at its present precision.

Events with four $b$ hadrons are 0.3\% of the $Z\to b\bar b$ events and change no bin by more than 0.5\%. The within-bin variation of $\bar\eps$ and $\bar D$ changes the amplitude by $-0.2\%$ and no bin by more than 1\% of the flux at its extrema. The migration between bins, with a direction resolution of $1.2^\circ$ against bins of $10^\circ$, leaves 85--90\% of the candidates in their true bin, and the true flux of the tagged hadrons binned in the reconstructed angle differs in amplitude from the same hadrons binned in the true angle by $(+0.2\pm0.2)\%$ with PYTHIA~8 and by $(-0.1\pm0.5)\%$ with the KK2f sample. The edge bins lose 5\% of their candidates to their inner neighbors because the efficiency rises steeply across them, and the efficiency measured in the same reconstructed bins absorbs it. The kinematic bias of the tagger (Sec.~\ref{sec:principle:residuals}) is estimated from the simulation at truth level. Each $b$ hadron is given a weight equal to the product of the efficiency and the dilution factor of the tagger at its own polar angle and scaled energy, $x_B=2E_B/\sqrt s$, where $E_B$ is the energy of the hadron. The amplitude of the weighted hadrons is then compared with that of all hadrons. It is higher by $(+1.9\pm0.2)\%$ with PYTHIA~8 and by $(+2.2\pm0.4)\%$ with the KK2f sample, because the tagger prefers hard $b$ hadrons, which carry the larger asymmetry. The direct comparison of the true flux of the tagged hadrons with that of all hadrons, which contains every other dependence of the selection, gives $(-1.2\pm1.6)\%$ and $(+0.4\pm2.8)\%$, consistent with it. The bias is $(+1.9\pm0.2)\%$, opposite in sign to the efficiency correlation. This first measurement is not binned in the energy of the $b$ hadron.

\begin{table}[h]
\caption{Uncertainties on the amplitude $A$ of the bottom flux, $\sin\theta\,\Fl(\theta)=A\sin2\theta$ (Eq.~\eqref{eq:sin2theta}), full-covariance fit over $30^\circ$--$150^\circ$. The calibration uncertainty is that of $\varepsilon$, $D$, and $\Delta$ determined from the data. The residuals are the possible biases of the method predicted from the PYTHIA~8 $Z\to b\bar b$ simulation, with their simulation-statistics errors. They are quoted, not applied, and the closure test bounds their sum. The systematic uncertainty is the change of the result when the full chain is run with the KK2f simulation in place of PYTHIA~8. The total combines the statistical, calibration, and systematic uncertainties in quadrature. No entry is given for the odd charge offset $\bar\Delta_{\rm o}$, which is controlled at track level by the charge-odd term of the momentum scale (Sec.~\ref{sec:edm4hep:momentum}).}
\label{tab:syst}
\centering
\begin{tabular}{lcc}
\hline
 & $\Delta A$ & $\Delta A/A$ \\
\hline
\multicolumn{3}{l}{\emph{Uncertainty of the measurement}} \\
statistical        & 0.0091 & 9.0\% \\
in-situ calibration & 0.0056 & 5.5\% \\
\hline
\multicolumn{3}{l}{\emph{Residuals predicted from the PYTHIA~8 simulation}} \\
efficiency correlation $C_{kk'}$ & & $-2.4\pm0.5\%$ \\
kinematic bias of the tagger     & & $+1.9\pm0.2\%$ \\
mistag covariance in $D$         & & $\phantom{+}0.0\pm1.1\%$ \\
angular migration                & & $+0.2\pm0.2\%$ \\
sum                              & & $-0.3\pm1.2\%$ \\
closure, reconstructed / truth   & & $1.021\pm0.048$ \\
\hline
\multicolumn{3}{l}{\emph{Systematic}} \\
PYTHIA~8 vs KK2f & 0.0016 & 1.6\% \\
\hline
Total & 0.0108 & 10.6\% \\
\hline
\end{tabular}
\end{table}

Table~\ref{tab:syst} collects the uncertainties and the residual effects. The residuals are predicted from the PYTHIA~8 simulation, which has a larger sample and current $b$-hadron decay tables. They are properties of the same candidate pairs and are correlated with one another, so they are added linearly rather than in quadrature, and their sum is compared with the closure test, which bounds the combined effect of all biases of the method that are present in the simulation. The odd charge offset, which the simulation does not contain, is controlled by the momentum calibration of Sec.~\ref{sec:edm4hep:momentum}. The calibration is measured on the data and its statistical precision is part of the quoted uncertainty; no further uncertainty is assigned to the method. Simulation enters the result only through the subtraction of the charm and light-quark candidates, and the systematic uncertainty is the change of the result when the full chain is run with the KK2f simulation in place of PYTHIA~8.

\section{Results}
\label{sec:results}

\begin{figure}[ht!]
\centering
\includegraphics[width=0.6\textwidth]{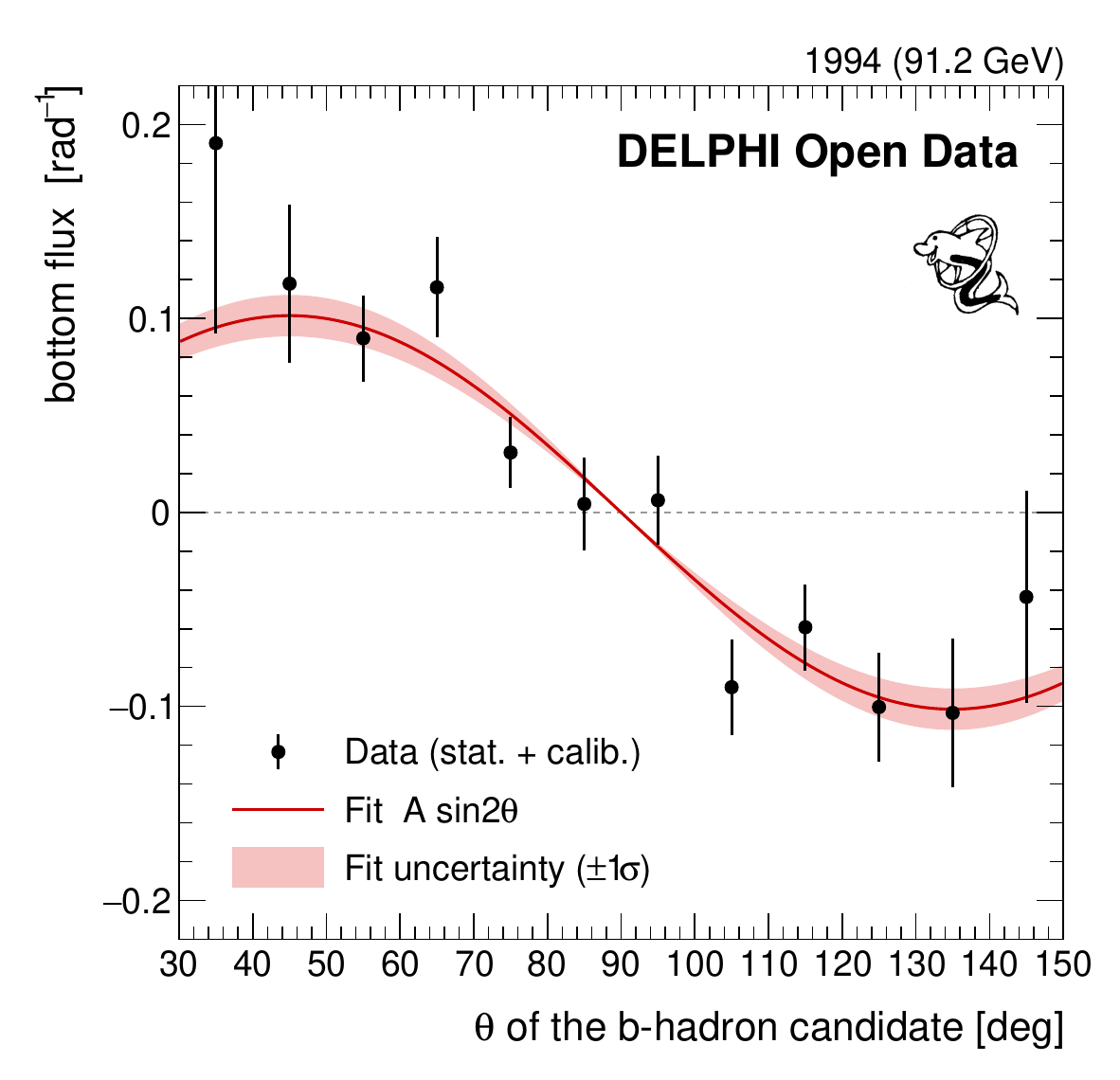}
\caption{Bottom flux measured in the 1994 data against the polar angle $\theta$ of the $b$-hadron candidate ($\theta=0$ along the $e^-$ beam), after the in-situ calibration of Sec.~\ref{sec:analysis:calibration}. The uncertainties are statistical and from the calibration. Line and band: fit of $A\sin2\theta$, Eq.~\eqref{eq:sin2theta}, and its $\pm1\sigma$ uncertainty.}
\label{fig:fluxdata}
\end{figure}

Figure~\ref{fig:fluxdata} shows the bottom flux measured in the 1994 data, after the calibration of Sec.~\ref{sec:analysis:calibration}. The modulation in $\sin2\theta$ of Eq.~\eqref{eq:sin2theta} is observed for the first time: the flux is positive in the forward hemisphere, where the $b$ quark goes, and negative in the backward one, it vanishes at $90^\circ$, and its extrema lie near $45^\circ$ and $135^\circ$. A fit of $A\sin2\theta$ to the twelve bins, with the full covariance from the bootstrap, determines the amplitude with a relative precision of 10.6\% in total: 9.0\% statistical and 5.5\% from the in-situ calibration, both from the 2000 bootstrap resamples of Sec.~\ref{sec:analysis:uncertainties}, and 1.6\% systematic. The systematic uncertainty is less than a quarter of the statistical one, and the residuals predicted from the simulation sum to $(-0.3\pm1.2)\%$ (Table~\ref{tab:syst}). The comparison of the amplitude with theory is presented in a companion paper. The measurement is limited by its statistics. A precision measurement, with a dedicated tagger and with the DELPHI data of all years at the $Z$ pole, is left for future work.

\section{Summary and Outlook}
\label{sec:summary}

The forward-backward asymmetry of $b$ quarks at the $Z$ pole has shown a longstanding tension of two to three standard deviations with the Standard Model. The LEP measurements estimated the direction of the $b$ quark from the thrust axis of the event, which makes the QCD correction to their result depend on the details of each analysis. In this paper we have measured a new observable sensitive to this asymmetry, the bottom flux. The bottom flux is the distribution of $b$ hadrons in polar angle, with each hadron counted as $+1$ or $-1$ depending on whether it contains a $b$ or a $\bar b$ quark. It uses the direction of each $b$ hadron instead of the direction of the quark, and its QCD correction does not depend on the event topology.

We have measured the bottom flux with the data recorded by the DELPHI experiment in 1994 at $\sqrt{s}=91.2$~GeV. Each $b$ hadron is identified by its displaced decay vertex, which also gives its direction. Whether it contains a $b$ or a $\bar b$ quark is inferred from the charge of the vertex or of a lepton from its decay. Not every $b$ hadron is found, and the inferred sign can be wrong. We have measured both effects from the data themselves, using events in which both $b$ hadrons are found, whose true signs are always opposite. The simulation is used only to subtract the small background from charm and light quarks and to estimate residual effects. Applied to simulated events, the full procedure recovers the true flux within its statistical precision.

The analysis uses the DELPHI open data and simulation in their final processing, which DELPHI prepared in its IDEA project as the data set for future analyses~\cite{Camporesi:2000xqf}. We have converted them to EDM4hep, the data format used in studies for future colliders, without reprocessing them. The converted data agree with the published performance of the DELPHI detector and show the shift of the track momentum scale that DELPHI itself had reported, which we correct. After the correction, the masses of known particles agree with their reference values, and positive and negative muons have consistent momenta in all regions of the detector, so that a detector asymmetry cannot imitate a signal. This measurement shows that the final DELPHI data set fulfills its purpose, supporting a complete physics analysis decades later, and that EDM4hep carries such an analysis, including the calibration of the detector response.

The measured flux has the expected shape: $b$ quarks are found more often in the direction of the electron beam, and $\bar b$ quarks in the opposite direction. From this shape we have determined the forward-backward asymmetry of the $b$ hadrons with a relative precision of 10.6\%. The measurement is limited by the size of the sample, and the systematic uncertainty is smaller than a quarter of the statistical one.

The statistical precision of this measurement depends on how many $b$ hadrons are found and how reliably their signs are identified. Taggers based on machine learning, which assign a sign to each decay vertex, could improve it substantially. The same method can be applied to the archived data of the other LEP experiments and of SLD, and at a future $Z$ factory.

\acknowledgments
We benefited from discussions with members of the DELPHI Collaboration involved in its data preservation and are especially grateful to Ulrich Schwickerath, whose sustained efforts to preserve the data and software, and to make them publicly available and usable, made this work possible. We thank Matteo Defranchis, Celia Lo, Michael Peskin, Huilin Qu, Manqi Ruan, and Ben Nachman for useful discussions.
The conversion framework released with this paper contributes to the effort to convert LEP data to EDM4hep~\cite{Fanini:2026mpx}.
We also thank Yi Chen, David d'Enterria, Brieuc Fran\c{c}ois, Yen-Jie Lee, Michele Selvaggi, and Birgit Stapf for their efforts in reviving interest in archived LEP data.
K.L. is supported by the U.S. Department of Energy under contract DE-AC02-06CH11357. I.M. is supported by the DOE Early Career Award DE-SC0025581, the Sloan Foundation, and the Simons Collaboration on Confinement and QCD Strings. S.Q. is supported in part by the U.S. CMS Operations Program. J.Z. is supported in part by the U.S. National Science Foundation under ID 2514008.

\section*{AI Use Disclosure}
The authors used the AI assistants Claude (Anthropic; versions Opus~5.5, Fable~5.1, and Fable~5.2) and Codex (OpenAI; versions 6~Astra and 5.5~Sol) in this work. Under the authors' direction, they assisted in writing the analysis software and the plotting scripts, in drafting and editing parts of the text, and in checking the manuscripts for consistency. The authors defined the method, reviewed the results, and revised all text, and they take full responsibility for the content.

\appendix

\section{Validation of the Converted DELPHI Data}
\label{app:validation}

We validate the converted data and simulation against the published performance of the DELPHI detector~\cite{DELPHI:1995dsm} by reproducing a set of its figures and tables with the 1994 data. We follow the paper's selection where given. Otherwise, we use cuts from other DELPHI publications, cited in each subsection. All comparisons are made at detector level, and the uncertainties of our results are statistical, unless stated otherwise. The studies use the 1994 hadronic data of Sec.~\ref{sec:edm4hep:samples}, leptonic data~\cite{DELPHI:OD:lolept94c2}, and DELPHI's simulation of $Z\to q\bar q$~\cite{DELPHI:OD:qqps_r94_c2}, $Z\to\mu^+\mu^-$~\cite{DELPHI:OD:dymu_r94_c2}, $Z\to\tau^+\tau^-$~\cite{DELPHI:OD:kora_r94_c2}, and $\gamma\gamma\to\mu^+\mu^-$~\cite{DELPHI:OD:ggmu_r94_c2}, all converted in the same way. As in the main analysis, runs are selected by the recorded state of the subdetectors (Sec.~\ref{sec:edm4hep:samples}). The integrated luminosities of the validation samples differ because each requires a different set of subdetectors to be operational. Our files come from DELPHI's final processing of the 1994 data, made in 1996 after the performance paper was completed. Where the reconstruction changed between processings, our results can differ from the published ones. We reproduce the momentum resolution (Sec.~\ref{app:validation:momentum}), the masses and widths of charm mesons (Sec.~\ref{app:validation:charm}), and the muon identification (Sec.~\ref{app:validation:muon}).

\subsection{Momentum Resolution}
\label{app:validation:momentum}

The momentum resolution is measured, as in Ref.~\cite{DELPHI:1995dsm}, with muons from $Z\to\mu^+\mu^-$ decays in the leptonic data. Events with exactly two tracks of opposite charge, both identified as muons by the standard tag of Sec.~\ref{app:validation:muon}, are selected. The two muons are required to have $\theta_{\rm acol}<0.15^\circ$, which removes radiative events. In the barrel, tracks with hits in the VD, ID, TPC, and OD and $42^\circ\leq\theta\leq138^\circ$ are used. In the forward region, tracks with hits in the VD and the FCB and $\theta\leq36^\circ$ or $\theta\geq144^\circ$ are used. The VD and the TPC are required to be operating at more than 99\% of their nominal, and the muon chambers at more than 90\%.

\begin{figure}[tbp]
\centering
\includegraphics[width=0.49\textwidth]{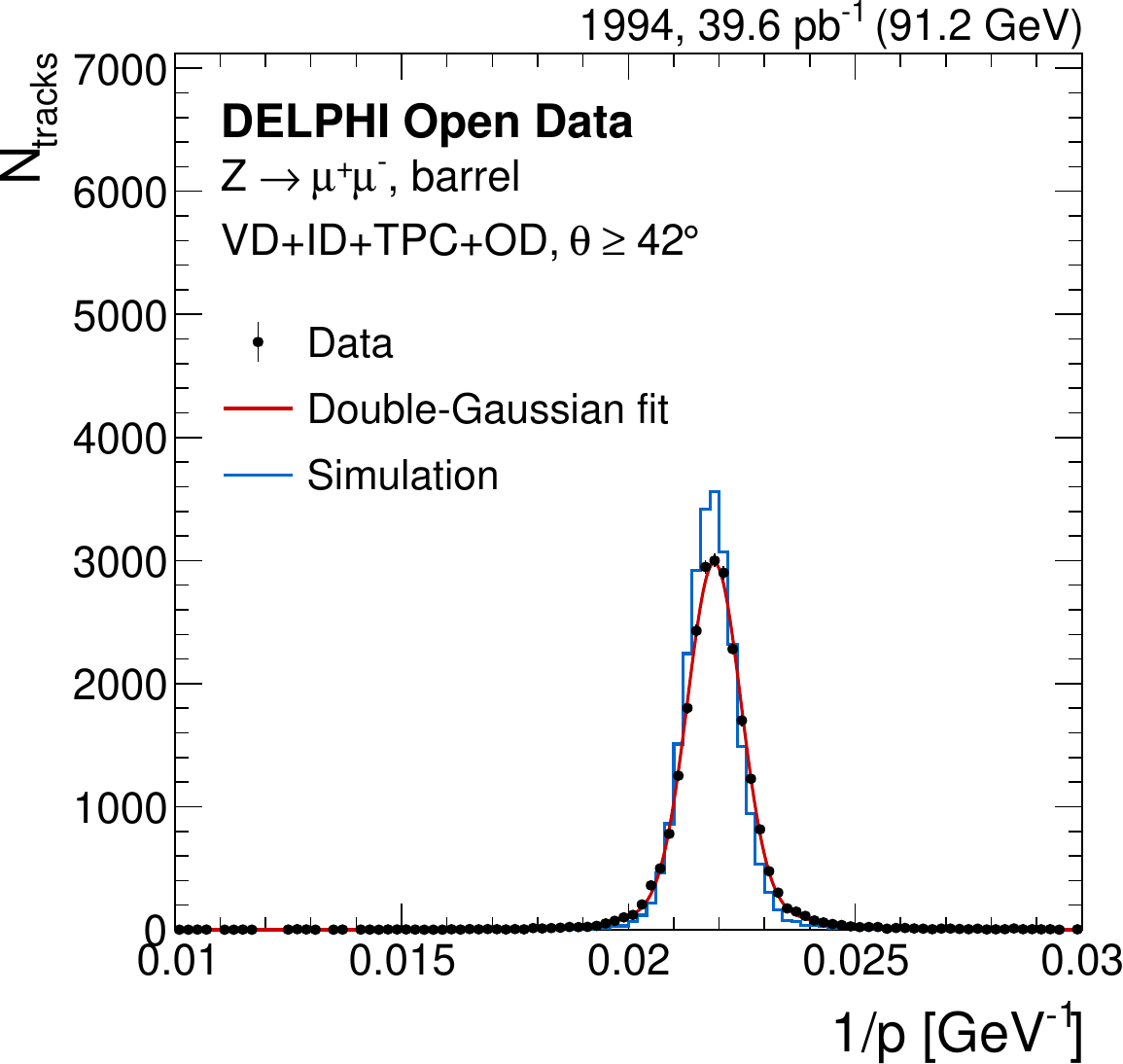}
\includegraphics[width=0.49\textwidth]{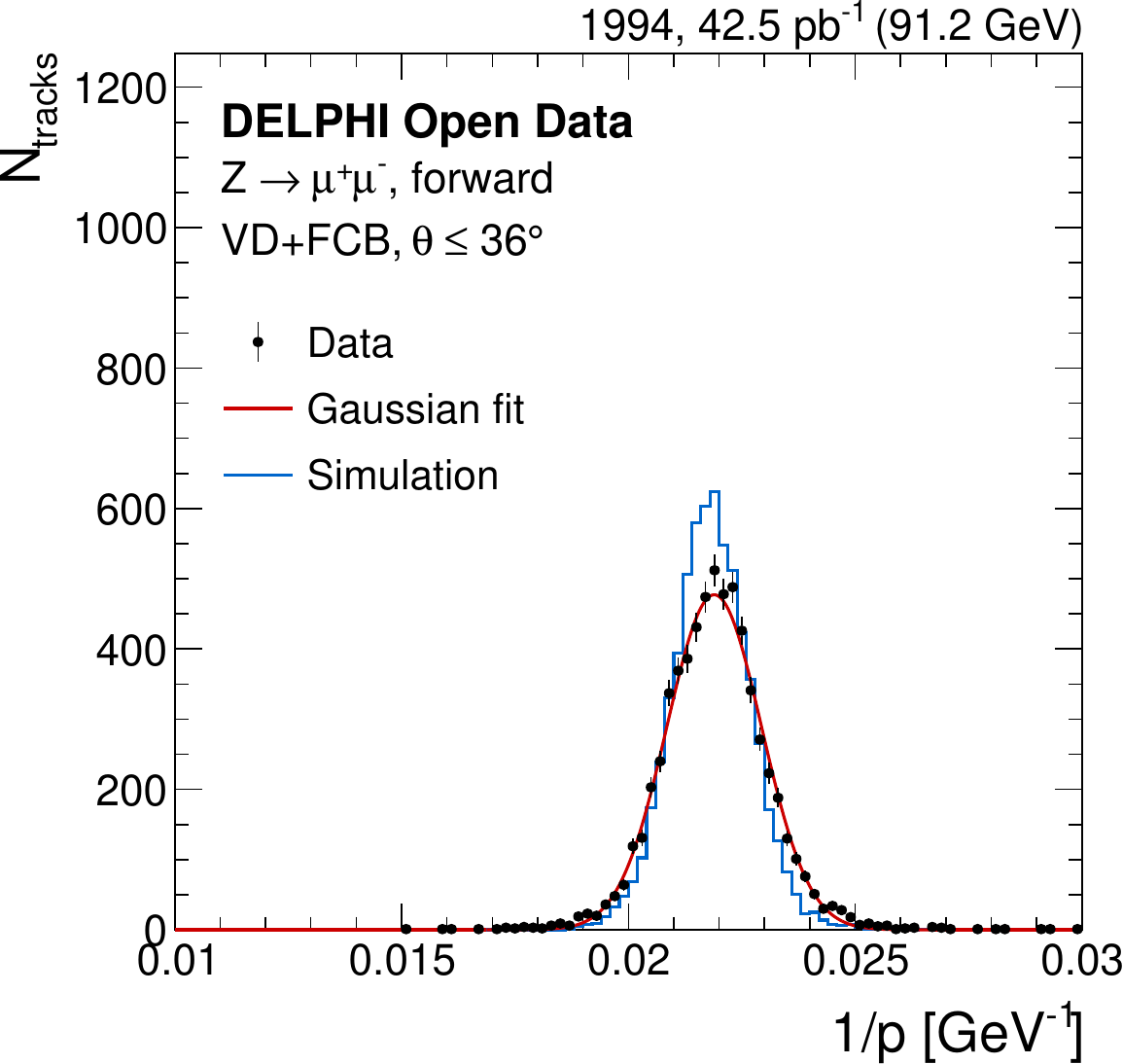}
\caption{Inverse momenta of muons in $Z\to\mu^+\mu^-$ events, for tracks in the barrel with hits in the VD, ID, TPC, and OD (left) and for tracks in the forward region with hits in the VD and the FCB (right). The data are shown as black markers and the simulation as a blue solid line. The simulation is normalized to the number of tracks in the data. The red solid line is a double-Gaussian fit (left) and a single-Gaussian fit (right) to the data, as in Ref.~\cite{DELPHI:1995dsm}.}
\label{fig:app:momentum}
\end{figure}

The left panel of Fig.~\ref{fig:app:momentum} shows the inverse momenta of the barrel tracks. As in Ref.~\cite{DELPHI:1995dsm}, the distribution is fitted with the sum of two Gaussians with a common mean. The narrower Gaussian has a width $\sigma(1/p)=(0.58\pm0.01)\times10^{-3}$~GeV$^{-1}$, compared with $0.57\times10^{-3}$~GeV$^{-1}$ in Ref.~\cite{DELPHI:1995dsm}. Varying the fit range changes it by about $0.01\times10^{-3}$~GeV$^{-1}$. The wider Gaussian describes the tails, with a width of $(1.52\pm0.05)\times10^{-3}$~GeV$^{-1}$, compared with $1.04\times10^{-3}$~GeV$^{-1}$. Its peak is $(6.7\pm0.7)\%$ of the total, compared with about 8\%. The simulation gives a narrower width of $(0.53\pm0.01)\times10^{-3}$~GeV$^{-1}$.

The right panel of Fig.~\ref{fig:app:momentum} shows the forward tracks. A single Gaussian gives $\sigma(1/p)=(1.04\pm0.01)\times10^{-3}$~GeV$^{-1}$, compared with $1.31\times10^{-3}$~GeV$^{-1}$ in Ref.~\cite{DELPHI:1995dsm}, and $(0.84\pm0.01)\times10^{-3}$~GeV$^{-1}$ in the simulation. Our resolution is better than the published one, which we attribute to the reprocessing of the data. Table~\ref{tab:app:momentum} compares the resolution for the track categories of Ref.~\cite{DELPHI:1995dsm}. The other barrel category and the forward category below $25^\circ$ agree with the published values within 5\%, while the forward category at $25$--$30^\circ$ is 22\% better. The category without the VD cannot be formed, since the VD was extended to the forward region in 1994.

\begin{table}[tbp]
\centering
\caption{Momentum resolution $\sigma(1/p)$ for $Z\to\mu^+\mu^-$ muons in the track categories of Ref.~\cite{DELPHI:1995dsm}, in units of $10^{-3}$~GeV$^{-1}$. The widths are from single-Gaussian fits, except in the first row, where the narrower Gaussian of the double-Gaussian fit is given. The category without the VD cannot be formed, as the VD also covered the forward region from 1994.}
\label{tab:app:momentum}
\begin{tabular}{llccc}
\toprule
$\theta$ & Detectors & Data & Simulation & Ref.~\cite{DELPHI:1995dsm} \\
\midrule
$\geq42^\circ$ & VD+ID+TPC+OD & $0.58\pm0.01$ & $0.53\pm0.01$ & 0.6 \\
$\geq42^\circ$ & ID+TPC+OD & -- & -- & 1.1 \\
$\geq42^\circ$ & VD+ID+TPC & $1.61\pm0.03$ & $1.51\pm0.02$ & 1.7 \\
$\leq36^\circ$ & VD+FCB & $1.04\pm0.01$ & $0.84\pm0.01$ & 1.3 \\
$25$--$30^\circ$ & FCB & $1.16\pm0.02$ & $0.97\pm0.01$ & 1.5 \\
$<25^\circ$ & FCB & $2.62\pm0.06$ & $2.67\pm0.08$ & 2.7 \\
\bottomrule
\end{tabular}
\end{table}

\subsection{Masses and Widths of Charm Mesons}
\label{app:validation:charm}

The masses and widths of $D$ mesons in hadronic events are compared in the channels $D^{*+}\to D^0(\to K^-\pi^+)\pi^+_s$, inclusive $D^0\to K^-\pi^+$, and $D^+\to K^-\pi^+\pi^+$, in two bins of $x_E=E_D/E_{\rm beam}$. The soft pion $\pi_s$ is required to have the charge opposite to the kaon. Charge-conjugate decays are included. The number and the total energy of the charged particles are denoted $N_{\rm ch}$ and $E_{\rm ch}$. The transverse and longitudinal impact parameters of a track are written $d_{xy}$ and $d_z$. We take the cuts from DELPHI's measurements of $D$-meson production and of the charm forward-backward asymmetry~\cite{DELPHI:1993gqe,DELPHI:1999ets,DELPHI:1999slw,DELPHI:1999qxi}, listed in Table~\ref{tab:app:charmcuts}. The $D^0$ and $D^+$ decay vertices are fitted from their tracks. A track with RICH information is accepted as a kaon unless the RICH gives it a standard or tight pion tag~\cite{DELPHI:1995dsm}. A track without RICH information needs at least the loose kaon tag of the TPC~\cite{DELPHI:1999ets}.

\begin{table}[tbp]
\centering
\caption{Selection of the $D$-meson candidates. The $\ell$ is the track length. The $d_{xy}^{\rm PV}$ and $d_z^{\rm PV}$ are the impact parameters with respect to the PV. The $d_{xy}^{\rm vtx}$ is the transverse distance of a track from the decay vertex. The $N_{\rm hit}^{\rm VD}$ and $N_{\rm hit}^{\rm OD}$ are the numbers of VD and OD hits of a track. The $\theta^*$ is the decay angle of the kaon in the $D^0$ rest frame. The $L$ is the decay length from the PV to the decay vertex, and $\sigma_L$ is its uncertainty. The $P(\chi^2)$ is the probability of the decay-vertex fit. Kaon identification (ID) is described in the text.}
\label{tab:app:charmcuts}
\setlength{\tabcolsep}{5pt}
\begin{tabular}{ll@{\hspace{2em}}lc}
\toprule
Object & \multicolumn{2}{l}{Cuts} & Ref. \\
\midrule
Event & $N_{\rm ch}\geq5$ & $E_{\rm ch}>0.12\sqrt s$ & \cite{DELPHI:1993gqe,DELPHI:1999ets,DELPHI:1999slw} \\
\cmidrule{1-4}
\multirow{3}{*}{Track} & $0.4<p<50$~GeV & $\sigma_p/p<1$ & \multirow{3}{*}{\cite{DELPHI:1999ets}} \\
 & $20^\circ<\theta<160^\circ$ & $\ell>30$~cm & \\
 & $|d_{xy}^{\rm PV}|<4$~cm & $|d_z^{\rm PV}|<10$~cm & \\
\cmidrule{1-4}
\multirow{2}{*}{All $D$} & $x_E>0.15$ & $p_K,p_\pi>1$~GeV & \cite{DELPHI:1993gqe,DELPHI:1999ets,DELPHI:1999qxi} \\
 & \multicolumn{2}{l}{$-0.5<L<2$~cm ($D^0$, $D^+$)} & \cite{DELPHI:1993gqe} \\
\cmidrule{1-4}
\multirow{3}{*}{$D^{*+}$} & \multicolumn{2}{l}{$N_{\rm hit}^{\rm VD}\geq1$ for $K$ and $\pi$} & \cite{DELPHI:1993gqe} \\
 & $0.4<p_{\pi_s}<3.5$~GeV & $|d_{xy}^{\rm PV}(\pi_s)|<3$~mm & \cite{DELPHI:1993gqe,DELPHI:1999qxi} \\
 & \multicolumn{2}{l}{$\cos\theta^*>-0.6$ ($x_E<0.25$), $-0.8$ ($x_E<0.5$), $-0.9$ ($x_E>0.5$)} & \cite{DELPHI:1993gqe} \\
\cmidrule{1-4}
\multirow{3}{*}{$D^0$ inclusive} & kaon ID & $P(\chi^2)>0.01$ & \multirow{3}{*}{\cite{DELPHI:1993gqe}} \\
 & $N_{\rm hit}^{\rm VD}\geq1$ for both tracks & $N_{\rm hit}^{\rm OD}\geq1$ for $\geq1$ track & \\
 & $L/\sigma_L>1$ & & \\
\cmidrule{1-4}
\multirow{4}{*}{$D^+$} & $p_K>2$~GeV & kaon ID for $x_E<0.55$ & \cite{DELPHI:1993gqe} \\
 & $N_{\rm hit}^{\rm VD}\geq1$ for $\geq2$ tracks & $d_{xy}^{\rm vtx}<100~\mu$m & \cite{DELPHI:1993gqe} \\
 & $P(\chi^2)>0.001$ & $L/\sigma_L>3$ & \cite{DELPHI:1993gqe,DELPHI:1999ets,DELPHI:1999qxi} \\
 & \multicolumn{2}{l}{$M_{K\pi\pi}-M_{K\pi}>200$~MeV for both pions} & \cite{DELPHI:1999ets,DELPHI:1999qxi} \\
\bottomrule
\end{tabular}
\end{table}

The mass distributions are fitted with the functions of Ref.~\cite{DELPHI:1993gqe}. The signal is a Gaussian in every channel. For the $D^{*+}$, the mass difference $\Delta M=M_{K\pi\pi_s}-M_{K\pi}$ is fitted for $1.79<M_{K\pi}<1.94$~GeV, where its background is described as a power law in $\Delta M-m_\pi$. The $D^0$ from $D^{*+}$ decays is fitted in $M_{K\pi}$ for $143.5<\Delta M<148.0$~MeV. The inclusive $D^0$ is fitted in $M_{K\pi}$ as well. Both have an exponential background. The $D^+$ is described with a linear background function. The fits start above 1.70~GeV, where partially reconstructed $D^0\to K^-\pi^+\pi^0$ decays do not contribute.

\begin{figure}[tbp]
\centering
\includegraphics[width=0.49\textwidth]{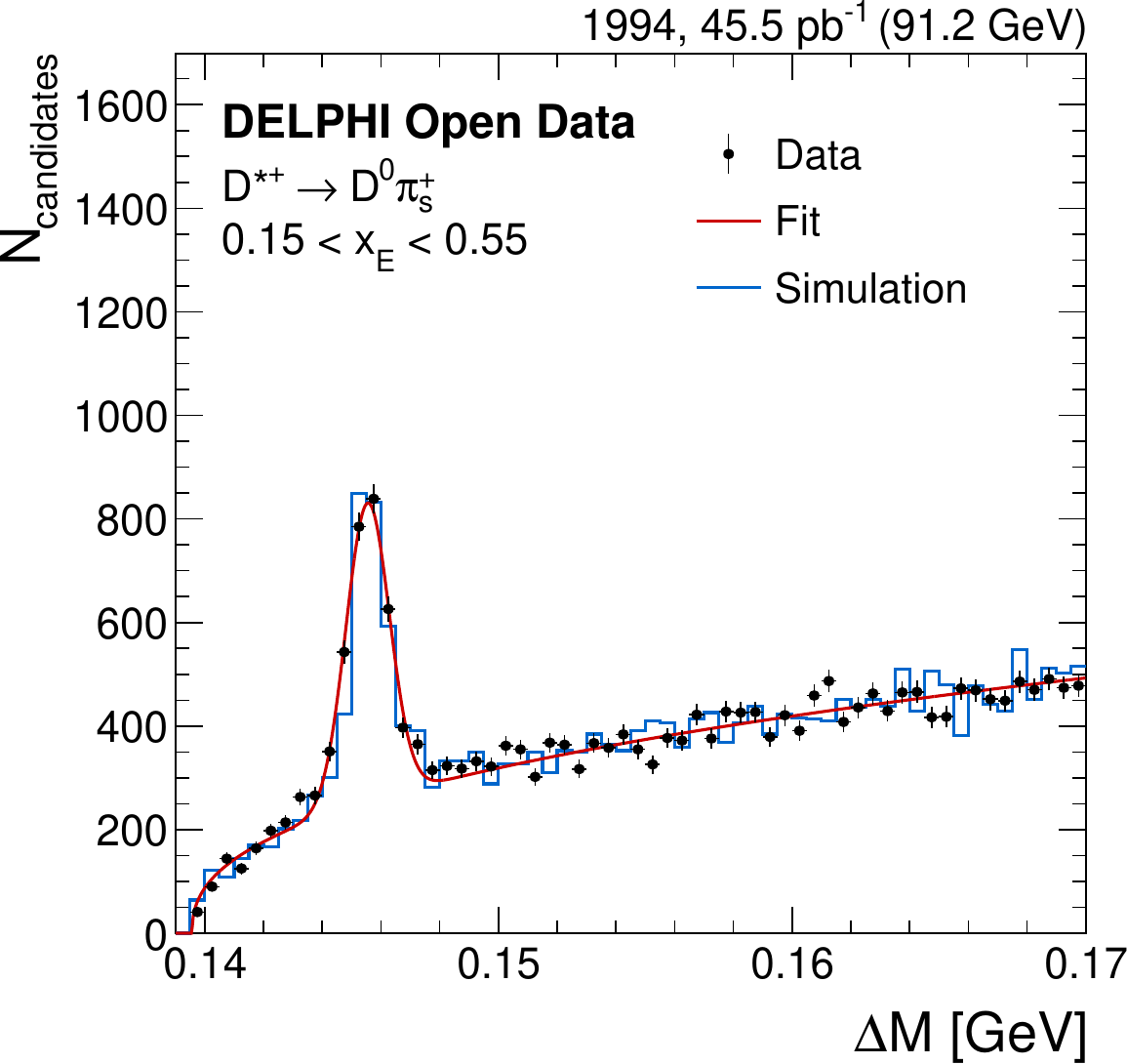}
\includegraphics[width=0.49\textwidth]{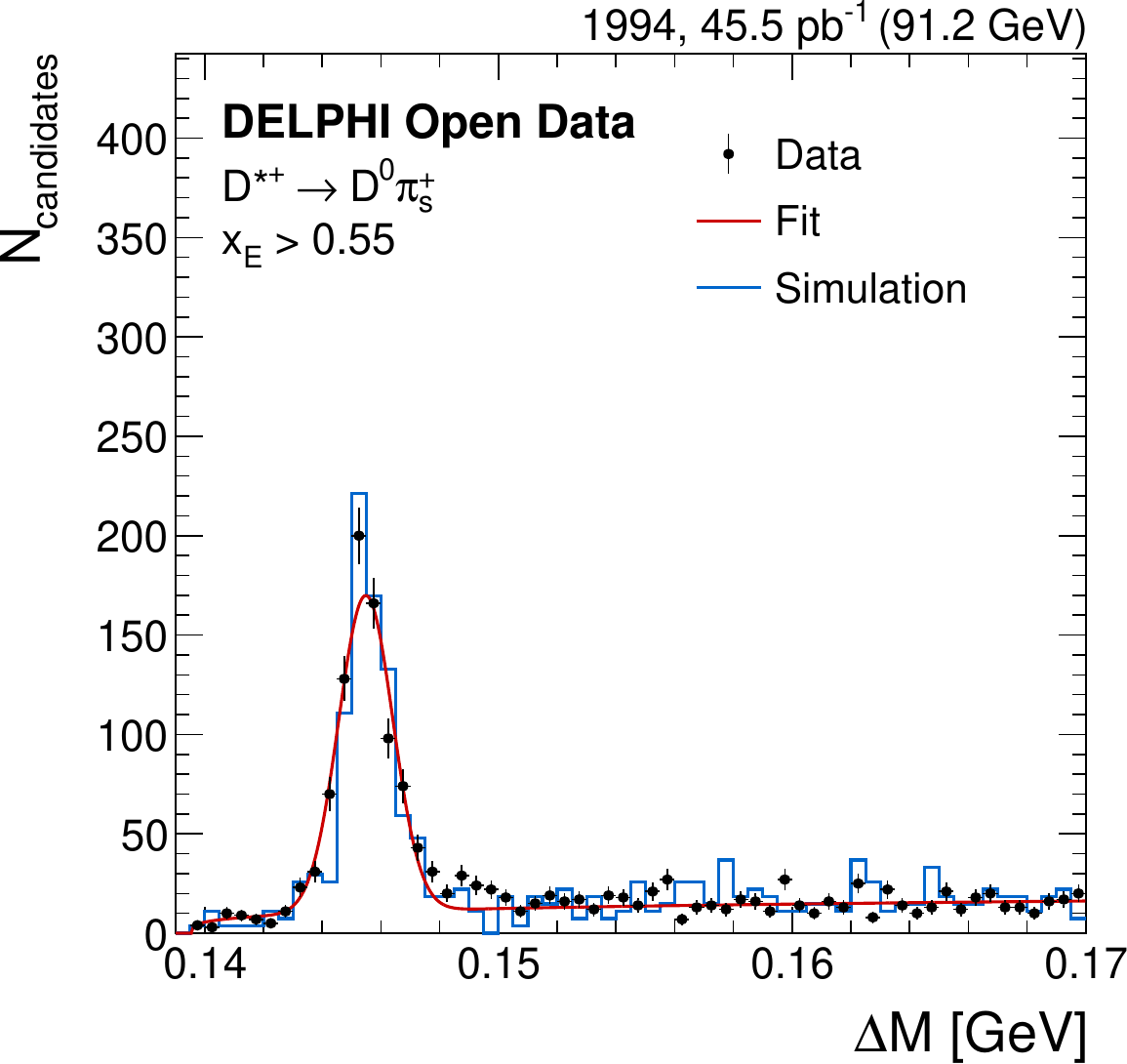}
\caption{Mass difference $\Delta M=M_{K\pi\pi_s}-M_{K\pi}$ of the $D^{*+}$ candidates for $0.15<x_E<0.55$ (left) and $x_E>0.55$ (right), with the stored momenta. The data are shown as black markers and the simulation as a blue solid line. The simulation is normalized to the number of candidates in the data. The red solid line is the fit to the data.}
\label{fig:app:dstar}
\end{figure}

Table~\ref{tab:app:charm} compares the fitted masses and widths with Table~5 of Ref.~\cite{DELPHI:1995dsm}. The widths of the $D^0$ and $D^+$ peaks agree, while the $\Delta M$ peak of the $D^{*+}$ is narrower (Fig.~\ref{fig:app:dstar}): $0.74\pm0.03$~MeV ($0.92\pm0.05$~MeV) at low (high) $x_E$, compared with $0.87\pm0.03$~MeV ($1.0\pm0.1$~MeV). The masses obtained with the stored momenta lie 2--7~MeV above those of Ref.~\cite{DELPHI:1995dsm}. After the momentum-scale correction of Sec.~\ref{sec:edm4hep:momentum}, they agree with the PDG values~\cite{ParticleDataGroup:2026mpi} within the uncertainties. The simulation agrees with the PDG values without the correction.

\begin{table}[tbp]
\centering
\caption{Masses $M$ and widths $\sigma$ of the $D$-meson peaks in MeV. For the $D^{*+}$, the width of the $\Delta M$ peak is given, and the tagged $D^0$ is the one from $D^{*+}$ decays. The data are shown with the stored and corrected momenta. For the corrected masses, the second uncertainty is the systematic uncertainty of the correction. The PDG masses are 1864.84~MeV ($D^0$) and 1869.66~MeV ($D^+$)~\cite{ParticleDataGroup:2026mpi}.}
\label{tab:app:charm}
\setlength{\tabcolsep}{5pt}
\small
\begin{tabular}{lllcccc}
\toprule
\multirow{2}{*}{Channel} & & \multirow{2}{*}{$x_E$} & \multicolumn{2}{c}{Data} & \multirow{2}{*}{Simulation} & \multirow{2}{*}{Ref.~\cite{DELPHI:1995dsm}} \\
\cmidrule{4-5}
 & & & Stored & Corrected & & \\
\midrule
\multirow{2}{*}{$D^{*+}$} & \multirow{2}{*}{$\sigma(\Delta M)$} & 0.15--0.55 & $0.74\pm0.03$ & $0.74\pm0.04$ & $0.62\pm0.03$ & $0.87\pm0.03$ \\
 & & $>0.55$ & $0.92\pm0.05$ & $0.91\pm0.04$ & $0.74\pm0.05$ & $1.0\pm0.1$ \\
\cmidrule{1-7}
\multirow{4}{*}{$D^0$, tagged} & \multirow{2}{*}{$M$} & 0.15--0.55 & $1868.5\pm1.0$ & $1864.0\pm0.9\pm2.3$ & $1864.2\pm1.2$ & $1862\pm1$ \\
 & & $>0.55$ & $1873.3\pm2.0$ & $1869.3\pm1.8\pm4.0$ & $1864.5\pm2.1$ & $1866\pm2$ \\
 & \multirow{2}{*}{$\sigma$} & 0.15--0.55 & $20.8\pm1.0$ & $20.7\pm0.9$ & $18.0\pm1.1$ & $22\pm1$ \\
 & & $>0.55$ & $30.7\pm2.0$ & $30.5\pm1.9$ & $17.7\pm1.8$ & $30\pm2$ \\
\cmidrule{1-7}
\multirow{4}{*}{$D^0$, inclusive} & \multirow{2}{*}{$M$} & 0.15--0.55 & $1868.6\pm0.8$ & $1864.3\pm0.8\pm2.6$ & $1865.8\pm1.2$ & $1864\pm1$ \\
 & & $>0.55$ & $1868.9\pm1.6$ & $1865.2\pm1.6\pm3.4$ & $1864.6\pm1.9$ & $1867\pm2$ \\
 & \multirow{2}{*}{$\sigma$} & 0.15--0.55 & $19.4\pm0.8$ & $19.2\pm0.8$ & $15.4\pm1.2$ & $22\pm1$ \\
 & & $>0.55$ & $24.0\pm1.6$ & $24.1\pm1.6$ & $21.6\pm1.8$ & $25\pm2$ \\
\cmidrule{1-7}
\multirow{4}{*}{$D^+$} & \multirow{2}{*}{$M$} & 0.15--0.55 & $1874.4\pm0.8$ & $1870.3\pm0.7\pm1.7$ & $1869.4\pm1.3$ & $1869\pm1$ \\
 & & $>0.55$ & $1873.8\pm1.1$ & $1870.0\pm1.1\pm2.6$ & $1872.3\pm1.4$ & $1868\pm1$ \\
 & \multirow{2}{*}{$\sigma$} & 0.15--0.55 & $15.7\pm0.8$ & $16.1\pm0.8$ & $15.1\pm1.3$ & $15\pm1$ \\
 & & $>0.55$ & $24.0\pm1.1$ & $24.1\pm1.1$ & $20.3\pm1.4$ & $21\pm2$ \\
\bottomrule
\end{tabular}
\end{table}

\subsection{Muon Identification}
\label{app:validation:muon}

Muons are identified by DELPHI's four tags: very loose, loose, standard, and tight~\cite{DELPHI:1995dsm}. Each tag fits the track, extrapolated through the hadron calorimeter, to the hits in the muon chambers. Hits with a large $\chi^2$ are removed and the fit is repeated, and the tag cuts on the $\chi^2$ of the final fit. As in Ref.~\cite{DELPHI:1995dsm}, the efficiency is measured for tracks with $p>3$~GeV. In the barrel, tracks with $53^\circ<\theta<127^\circ$ are used, excluding the gap between the two halves of the muon chambers at $88.5^\circ<\theta<91.5^\circ$. In the end caps, tracks with $20^\circ<\theta<42^\circ$ or $138^\circ<\theta<160^\circ$ are used.

The efficiencies are measured with muons selected without the muon chambers, in three samples: $Z\to\mu^+\mu^-$ for muons of about 45~GeV, $Z\to\tau^+\tau^-$ for 3--40~GeV, and $\gamma\gamma\to\mu^+\mu^-$ for low momenta. Their selections are listed in Table~\ref{tab:app:muoncuts}, where the impact parameters are taken with respect to the interaction point. In $Z\to\mu^+\mu^-$ and $\gamma\gamma\to\mu^+\mu^-$ events, the efficiency is measured on the muon other than the one tagged as a tight muon. The $Z\to\tau^+\tau^-$ events are divided into two hemispheres by the thrust axis. The efficiency is measured on the track of a single-track hemisphere with $3<p<40$~GeV, identified as a muon by the calorimeters. This track deposits less than 1~GeV in the electromagnetic calorimeters and less than 5~GeV in the hadron calorimeter. It deposits energy in at least three of the four layers of the hadron calorimeter, including the outermost. Events with a single track in each hemisphere must in addition have $\theta_{\rm acol}>0.5^\circ$, $|d_{z,a}-d_{z,b}|<3$~cm, and $|\sum\vec p_T|>0.4$~GeV, where the sum runs over the tracks and the electromagnetic clusters.

\begin{table}[tbp]
\centering
\caption{Selection of the muon and pion samples. The $a$ and $b$ label the leading tracks of the two hemispheres, and $E_{a,b}$ are the electromagnetic energies within $30^\circ$ of them. The $\theta_{\rm iso}$ is the smallest angle between tracks in opposite hemispheres. The $E_{\rm vis}$ is the sum of the track momenta and the electromagnetic energy.}
\label{tab:app:muoncuts}
\setlength{\tabcolsep}{5pt}
\begin{tabular}{ll@{\hspace{2em}}l}
\toprule
Sample & \multicolumn{2}{l}{Cuts} \\
\midrule
\multirow{2}{*}{$Z\to\mu^+\mu^-$} & $p>40$~GeV, $20^\circ<\theta<160^\circ$ & $\theta_{\rm acol}<10^\circ$ \\
 & $|d_{xy}|<1.5$~cm & $|d_z|<4.5$~cm \\
\cmidrule{1-3}
\multirow{4}{*}{$Z\to\tau^+\tau^-$} & \multicolumn{2}{l}{$2\leq N_{\rm ch}\leq6$ ($|d_{xy}|<5$~cm and $|d_z|<10$~cm)} \\
 & $43^\circ<\theta<137^\circ$ for $a$ or $b$ & $\theta_{\rm iso}\geq160^\circ$ \\
 & $E_{\rm vis}>8$~GeV & $\sqrt{p_a^2+p_b^2}/E_{\rm beam}<1$ \\
 & $\sqrt{E_a^2+E_b^2}/E_{\rm beam}<1$ & $|d_{xy}|<1.5$~cm, $|d_z|<4.5$~cm \\
\cmidrule{1-3}
\multirow{3}{*}{$\gamma\gamma\to\mu^+\mu^-$} & $N_{\rm ch}=2$, opposite charges & $p>3$~GeV, $\theta_{\rm acol}>5^\circ$ \\
 & $E_{\rm vis}<20$~GeV & $|\vec p_{T,1}+\vec p_{T,2}|<2$~GeV \\
 & $|d_{xy}|<1.5$~cm & $|d_{z,1}-d_{z,2}|<3$~cm \\
\bottomrule
\end{tabular}
\end{table}

\begin{figure}[tbp]
\centering
\includegraphics[width=0.49\textwidth]{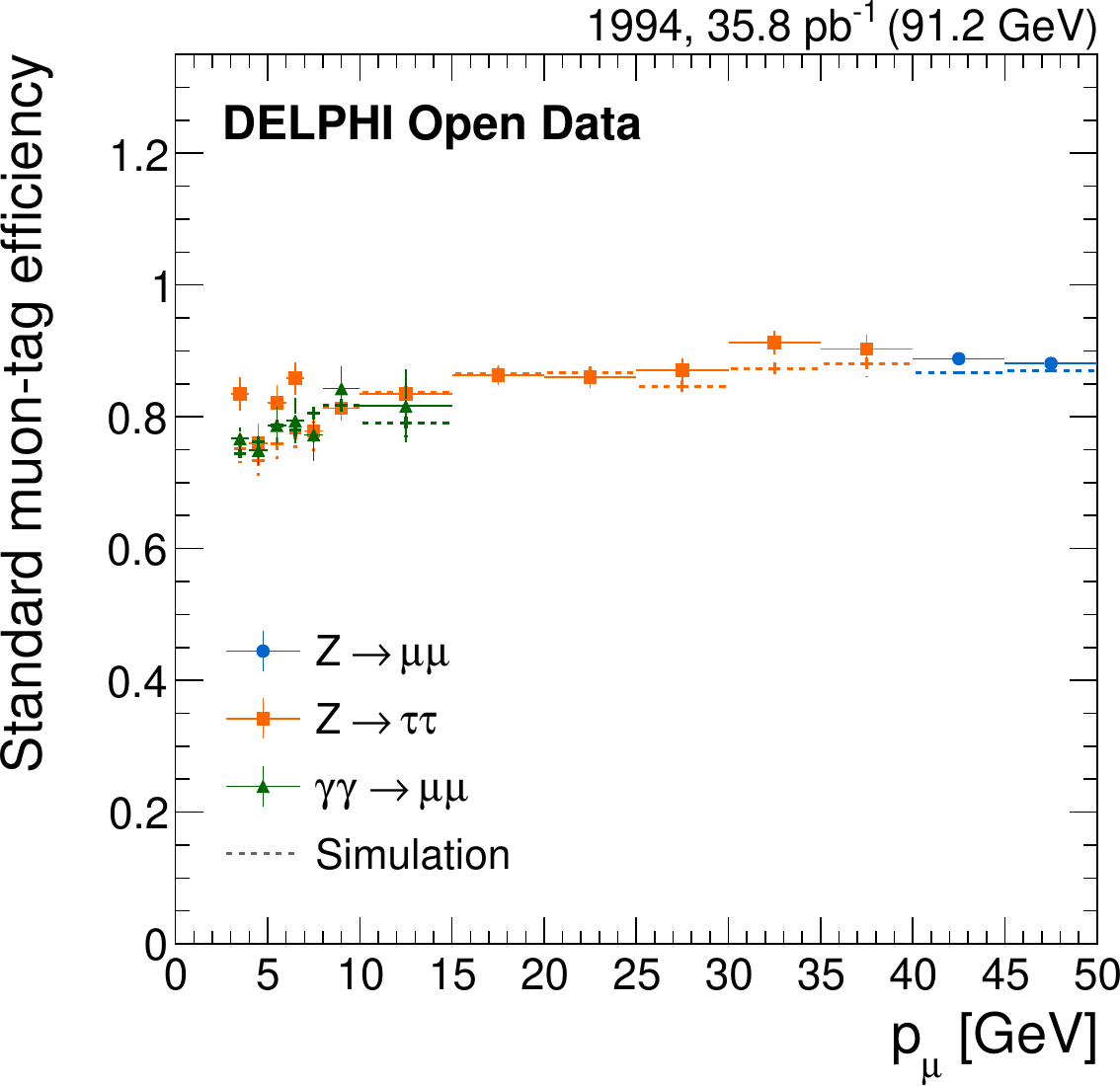}
\includegraphics[width=0.49\textwidth]{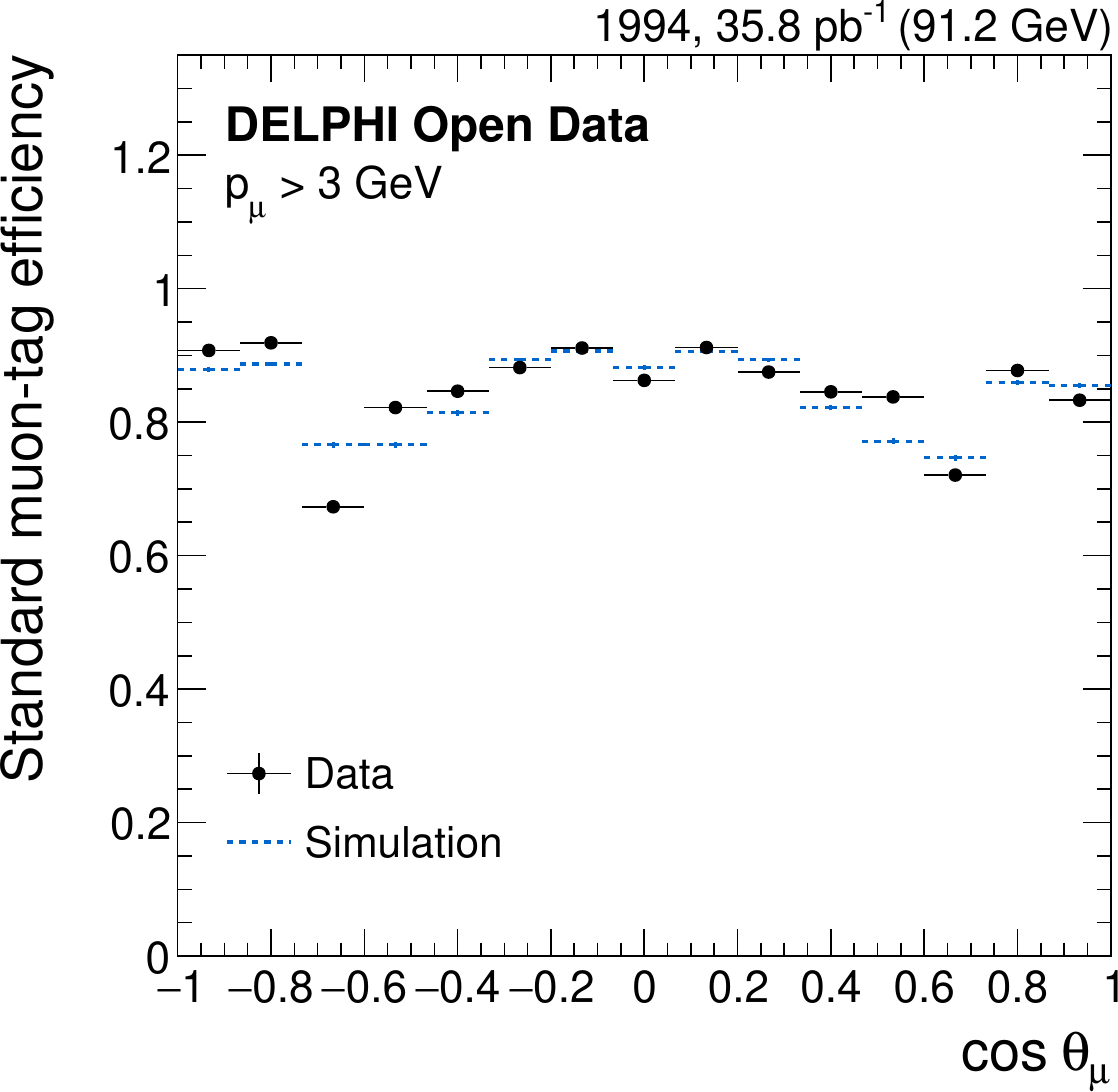}
\caption{Efficiency of the standard muon tag as a function of the momentum (left) and of $\cos\theta$ for $p>3$~GeV (right). The data (simulation) are shown as markers (dashed lines). The $Z\to\mu^+\mu^-$, $Z\to\tau^+\tau^-$, and $\gamma\gamma\to\mu^+\mu^-$ samples are illustrated as blue, orange, and green, respectively. The three samples are combined in the right panel.}
\label{fig:app:muon}
\end{figure}

The misidentification probability is measured on pions from $\tau\to3\pi\nu$ decays, as in Ref.~\cite{DELPHI:1995dsm}. The pions are the tracks with $p>3$~GeV in hemispheres with three tracks and a total charge of $\pm1$ in the $Z\to\tau^+\tau^-$ sample. Runs are used when the TPC, HPC, and hadron calorimeter were operating at more than 99\% of their nominal. The VD, FEMC, and muon chambers are required at more than 90\%. Table~\ref{tab:app:muon} compares the efficiencies and misidentification probabilities with Table~9 of Ref.~\cite{DELPHI:1995dsm}. The left panel of Fig.~\ref{fig:app:muon} shows the efficiency of the standard tag as a function of the momentum for the three samples, and the right panel shows it as a function of $\cos\theta$.

\begin{table}[tbp]
\centering
\caption{Efficiency and misidentification probability of the four muon tags in percent. The efficiencies are measured on the three muon samples combined, and the misidentification probabilities on pions from $\tau\to3\pi\nu$ decays. The uncertainties are statistical.}
\label{tab:app:muon}
\begin{tabular}{lcccccc}
\toprule
\multirow{2}{*}{Tag} & \multicolumn{3}{c}{Efficiency} & \multicolumn{3}{c}{Misidentification} \\
\cmidrule(lr){2-4}\cmidrule(lr){5-7}
 & Data & Simulation & Ref.~\cite{DELPHI:1995dsm} & Data & Simulation & Ref.~\cite{DELPHI:1995dsm} \\
\midrule
Very loose & $96.1\pm0.1$ & $98.0\pm0.1$ & $95.9\pm0.1$ & $4.54\pm0.18$ & $3.08\pm0.11$ & $5.4\pm0.2$ \\
Loose      & $94.5\pm0.1$ & $94.5\pm0.1$ & $94.8\pm0.1$ & $1.34\pm0.10$ & $0.82\pm0.06$ & $1.5\pm0.1$ \\
Standard   & $87.5\pm0.1$ & $86.0\pm0.1$ & $86.1\pm0.2$ & $0.73\pm0.07$ & $0.49\pm0.04$ & $0.7\pm0.1$ \\
Tight      & $75.0\pm0.2$ & $75.4\pm0.1$ & $76.0\pm0.2$ & $0.49\pm0.06$ & $0.32\pm0.04$ & $0.4\pm0.1$ \\
\bottomrule
\end{tabular}
\end{table}

\clearpage
\bibliographystyle{cms_unsrt}
\bibliography{paper/refs}

\end{document}

%% file: notes/notation.tex
\newcommand{\Ab}{\mathcal{A}_b}                 %
\newcommand{\Ae}{\mathcal{A}_e}                 %
\newcommand{\AFBz}{A^{0,b}_{\rm FB}}            %
\newcommand{\sbb}{\sigma_{b\bar b}}             %
\newcommand{\sbbz}{\sigma^{(0)}_{b\bar b}}      %
\newcommand{\stot}{\sigma_{\rm tot}}            %

\newcommand{\Cqcd}{C_{\rm QCD}}                 %
\newcommand{\LQCD}{\Lambda_{\rm QCD}}

\newcommand{\Bn}{\mathcal{B}}                   %
\newcommand{\Qdet}{\mathcal{Q}_{\Bn}}            %
\newcommand{\Fl}{\mathcal{F}}                   %
\newcommand{\Fh}{\hat{\mathcal{F}}}             %
\newcommand{\nh}{\hat{n}}                       %
\newcommand{\Nhad}{N_{\rm had}}
\newcommand{\Nbb}{N_{b\bar b}}

\newcommand{\eps}{\varepsilon}                  %
\newcommand{\fSS}{f_{SS}}                       %
\newcommand{\Cov}{\mathrm{Cov}}